\documentclass{IEEEtran}
\IEEEoverridecommandlockouts
\usepackage{cite}
\usepackage{amsmath,amssymb,amsfonts}
\usepackage{algorithmic}
\usepackage{graphicx}
\usepackage{textcomp}
\usepackage[table]{xcolor}
\usepackage{setspace}
\def\BibTeX{{\rm B\kern-.05em{\sc i\kern-.025em b}\kern-.08em
		T\kern-.1667em\lower.7ex\hbox{E}\kern-.125emX}}
\usepackage{subfigure}
\newtheorem{my_theorem}{Theorem}

\newtheorem{my_lemma}{Lemma}
\newtheorem{my_corollary}{Corollary}
\newtheorem{my_proposition}{Proposition}

\title{Performance Analysis of Dual-Hop Relaying for THz-RF Wireless Link with Asymmetrical Fading}
\author{Pranay Bhardwaj,~\IEEEmembership{Graduate Student Member,~IEEE} and S.~M.~ Zafaruddin,~\IEEEmembership{Senior Member,~IEEE}
	\thanks{A conference version of the paper  is accepted for presentation at the 2021 IEEE 93rd Vehicular Technology Conference (VTC2021-Spring), Helsinki, Finland, 25-28 April 2021 \cite{pranay2021}.}
	\thanks{ This work was supported in part by the Science and
		Engineering Research Board (SERB), Department of Science and Technology
		(DST), Government of India, under Start-up Research Grant SRG/2019/002345.}
	\thanks{Pranay Bhardwaj (p20200026@pilani.bits-pilani.ac.in) and S.~M.~Zafaruddin (syed.zafaruddin@pilani.bits-pilani.ac.in)  are  with  the Department of Electrical and Electronics Engineering, Birla Institute of Technology and Science, Pilani, Pilani-333031, Rajasthan, India.}
	
	\thanks{}}

\usepackage{enumitem}
\begin{document}
	\maketitle 

\begin{abstract} 
	Terahertz (THz) frequency bands can be promising for data transmissions between the core  network and  access points (AP)  for next-generation wireless systems. In this paper, we analyze the performance of a dual-hop  THz-RF wireless system where an AP facilitates data transmission between a core network  and  user equipment (UE). We consider a generalized model for the end-to-end  channel with independent and not identically distributed  (i.ni.d.)  fading  model  for  THz and RF links using the $\alpha$-$\mu$ distribution, the THz link with pointing errors, and asymmetrical relay position. We derive a closed-form expression of the cumulative distribution function (CDF) of the end-to-end signal to noise ratio (SNR) for the  THz-RF link, which is valid for continuous values of $\mu$ for a generalized performance analysis over THz fading channels. Using the derived CDF, we analyze the performance of THz-RF relayed system using decode-and-forward (DF) protocol by deriving analytical expressions of diversity order, moments of SNR, ergodic capacity, and average BER  in terms of system parameters. We also analyze the considered system with  an  i.i.d. model, and  develop simplified performance to provide insight on the system  behavior analytically  under various practically relevant scenarios. Simulation and numerical analysis show significant effect of fading parameters  of the THz link  and a nominal  effect of normalized beam-width on the performance of the relay-assisted THz-RF system.
\end{abstract}		
\begin{IEEEkeywords}
	 Bit error rate, Decode and forward, Diversity order, Ergodic capacity, Outage Probability, Performance analysis, Pointing errors, Relaying, Signal to noise ratio, Terahertz.
\end{IEEEkeywords}

\section{Introduction}
Network densification is a potential technology to support the widespread proliferation of high data rate applications for a large number of devices \cite{Bhushan2014densification,Liu2017densification,Dulaimi2017densification}. This densification can be realized by adding more cell sites, including radio access networks (RAN), macro sites, small cell deployments, the cell-free architecture, and the use of novel Terahertz (THz) spectrum.  The THz can provide tremendously high, unlicensed bandwidth which can be central to ubiquitous wireless communications in beyond-5G or sixth-generation (6G) networks \cite{Koenig_2013_nature,Elayan_2019,faisal2020}. Generally, the devices are connected over the radio frequency (RF) to a nearby access point (AP), which transports data to the core network through a high-speed back-haul link. The wire-line back-haul may consist of digital subscriber lines (DSL) and optical fiber. However, wire-line links might not be feasible in some adverse situations \cite{Wang2015}.  In contrast to the wireless back-haul link over the RF frequencies, THz wireless systems can be a promising alternative for high data rate transmission between the core  and  AP.  Until the transition to a complete THz system, the communication link between a user and the AP will continue on the conventional low-frequency RF transmissions. It is desirable to evaluate the performance of a  heterogeneous THz-RF link  for the next generation of wireless networks.

The THz link suffers from pointing errors when there is a misalignment between transmitter and receiver antenna beams at higher frequencies besides the signal fading and significantly higher path loss.  Recently, the use of THz for wireless communications is gaining research interests  \cite{Wang2014,Boulogeorgos_Analytical,Boulogeorgos_Error,Jornet_2011,  Priebe_2011,Kim2015,Kokkoniemi_2018,Wu2020,Sarieddeen2019,Cheng_2020,Olutayo_2020,Bian2021,boronin2015,Petrov2015,zhang2016,Elayan2018noisemodel,Boluda_2017,KOKKONIEMI2020,Rong_2017,Abbasi_2017,Boulogeorgos_20020_THz_Relaying,Mir2020}.  The authors in \cite{Wang2014} proposed a prototype of  a wireless local area network at $0.34$ THz frequency band.  The authors in \cite{Boulogeorgos_Analytical} have analyzed the ergodic capacity and outage probability for a single-link  THz system by deriving distribution functions for the combined  effect of  pointing errors\cite{Farid2007} and $\alpha$-$\mu$ fading \cite{Yacoub_alpha_mu}.  The $\alpha$-$\mu $ is a generalized model that includes other fading models as a particular case, and it is well studied in the RF context.  Recently, the authors in  \cite{Boulogeorgos_Error} have derived the average bit-error-rate (BER) and the outage probability of a mixed  THz-RF link with decode and forward (DF) relaying. However, derived distribution functions  are valid only for integer values of the fading parameter $\mu$. Further, the short term fading  is assumed to be independent and   identically distributed (i.i.d)  with the same parameters $\alpha$  and $\mu$ for both the THz and RF channels. The i.i.d. model might not be possible considering two  technologies operating over the different spectrum.  Moreover, there is no analysis available in the open literature for the average signal to noise ratio (SNR) and ergodic rate   performance for the THz-RF relaying.  Performance bounds on the outage probability,  average SNR,  ergodic rate, and average BER are  desirable for real-time tuning of system parameters for  efficient deployment of the THz-RF system.  

In this paper, we analyze the performance of a THz-RF link with DF relaying for data transmission between the central processing unit of a core network and a user through an AP in a  wireless network  over $\alpha$-$\mu$ fading channels. The major contributions of the paper are as follows:\vspace{-0.1mm}
\begin{itemize}[leftmargin=*]
	\item We consider a generalized model for the end-to-end  channel with independent and not identically distributed  (i.ni.d.)  fading  model  for  THz and RF links, the THz link with pointing errors, and asymmetrical relay position between the source and destination.
	\item  We derive a closed-form expression of the cumulative distribution function (CDF) for the SNR of the  THz link using the combined effect of $\alpha$-$\mu$ fading  and pointing errors. The derived CDF is also valid for non-integer values of $\mu$ for a generalized performance analysis over THz fading channels. 
	\item We analyze the performance of the THz-RF relayed system by deriving analytical expressions of diversity order, moments of SNR, ergodic capacity, and average BER  in terms of system parameters. The derived expressions are expressed in well-known mathematical functions, and can be evaluated using standard computational software. 
	\item Considering the i.i.d. model  of the short term fading using the same $\alpha$ and $\mu$ parameters for  THz and RF links,  we derive analytical expressions on the average SNR, ergodic capacity, and average BER of the  relay-assisted system. The  i.i.d. model simplifies the derived analytical expressions and provide insight on the system performance analytically.  The average SNR and ergodic capacity performance of the THz-RF system with the i.i.d. model  is accepted for presentation in the conference version of this paper \cite{pranay2021}.
	\item We validate the analytical results using numerical and Monte Carlo simulations. The computer simulations show a significant effect of fading parameters  of the THz link  and a nominal  effect of normalized beam-width on the performance of the relay-assisted THz-RF system.	
\end{itemize}


\subsection{Related Work}
The THz spectrum can provide tremendously high unlicensed bandwidth,  which can be a catalyst for next-generation  wireless technologies.  However, the path-loss in the THz band is higher due to molecular absorption of the signal at extremely small wavelengths \cite{Kim2015,Kokkoniemi_2018, Wu2020,Sarieddeen2019}. In \cite{Kim2015}, the authors  have presented an experimental characterization of  the THz channel ($300$-$320$ \mbox{GHz}).  The effect of scattering and absorption losses on THz wavelengths in the absence of white noise  were  examined  in \cite{Kokkoniemi_2018}. Signal prorogation in the indoor environment with the blockage effects by the walls and human bodies was analyzed in   \cite{Wu2020}. A spatial modulation technique to mitigate the path-loss at THz frequencies using  the properties of densely packed configurable arrays of nano-antennas  was presented in \cite{Sarieddeen2019}. 

In addition to the path loss, the THz transmissions undergo signal fading due to the multi-path propagation \cite{Boulogeorgos_Analytical,Boulogeorgos_Error,Jornet_2011,Priebe_2011,Cheng_2020,Kim2015,Kokkoniemi_2018,Olutayo_2020,Wu2020,Bian2021}. The authors in \cite{Jornet_2011} used the radiative transfer theory with  molecular absorption to present a propagation channel model for short-distance THz transmissions.  In \cite{Priebe_2011}, an experimental channel model in an indoor environment at $300$ \mbox{GHz} frequency for  line-of-sight (LOS) and non-LOS scenarios was proposed. The authors in \cite{Cheng_2020} have considered $ m $-Nakagami fading to model a 4$\times$4 THz multiple-input  multiple-output (MIMO) system.  A shadowed Beaulieu-Xie (BX) fading model was suggested in \cite{Olutayo_2020}. The authors in \cite{Bian2021} used advanced channel characteristics such as spherical wavefront, time-variant velocities, and space-time frequencies  to model  a three-dimensional non-stationary channel  for millimeter-wave and THz transmissions. Recently, the generalized $\alpha$-$\mu$  and fluctuating two ray (FTR) fading models are employed for THz channels due to their tractability in performance analysis \cite{Boulogeorgos_Analytical} \cite{RIS_THz_HW_Impaiment}.

As is for any communication systems, THz systems are  impacted by  noise,  interference \cite{boronin2015,Petrov2015,zhang2016,Elayan2018noisemodel}, and   pointing errors \cite{Boulogeorgos_Error,Boulogeorgos_Analytical,Boluda_2017,KOKKONIEMI2020}. The effect of  pointing errors at higher frequencies  is inevitable even with highly directional signal beams.  The authors in \cite{KOKKONIEMI2020} improvised antenna gains  for THz frequencies under the beam misalignment error  due to the movement of  antennas.  Significant research has recently been carried out to model the noise and interference for THz systems \cite{boronin2015,Petrov2015,zhang2016,Elayan2018noisemodel}. A simple and novel noise model and parameters fitting algorithm for THz band considering the molecular noise behavior  has been studied in \cite{boronin2015}. An analytical model for interference  in dense randomly deployed THz network was presented in \cite{Petrov2015}. The authors in \cite{zhang2016} proposed a model for channel noise inside human tissues at the THz band considering the molecular absorption.  The authors in \cite{Elayan2018noisemodel} presented a noise model for intra-body system including Johnson-Nyquist, Black-body, and Doppler-shift induced noise at THz frequencies.

Relay-assisted communication is a potential technique to deal with channel fading in wireless systems \cite{Nosratinia2004, Li2012,Bjornson_2013}. The use of relaying at THz frequencies has recently been discussed in  \cite{Rong_2017,Abbasi_2017,Boulogeorgos_20020_THz_Relaying,Mir2020}. Relaying for nano-scale THz transmissions was studied in \cite{Rong_2017,Abbasi_2017}. The outage probability of a THz-THz relaying scheme with MIMO is presented in \cite{Boulogeorgos_20020_THz_Relaying}. The authors in \cite{Mir2020} proposed a simplified hybrid precoding design for THz-MIMO communication system consisting of a two way amplify-forward (AF) relay with orthogonal frequency division multiplexing (OFDM). To this end, we should note that the free-space optical (FSO)  is another high-frequency wireless system, which can be impacted by pointing errors.   Hybrid RF/FSO systems have been studied in  \cite{Lee2011, Ansari2013, Samimi2013, assym_rf_fso2015, series_hybrid_m_channel2015, dual_hop_rf_fso_turb2016, Bag2018,Zhang2020}.   A dual-hop RF-FSO relay system over the asymmetric links was studied in \cite{Lee2011}. The BER performance and the capacity analysis of an AF-based dual-hop mixed RF–FSO were presented in \cite{Ansari2013}.  The authors in \cite{Samimi2013} have investigated the end-to-end outage performance, where the RF and FSO links were  modeled as Rayleigh and $M$-distributed, respectively. The authors in  \cite{assym_rf_fso2015, series_hybrid_m_channel2015, dual_hop_rf_fso_turb2016} have analyzed the FSO performance under the  turbulence and pointing errors using a dual-hop transmission with a single-relay without direct transmissions. Recently, a dual-hop relaying combined with mmWave  and FSO  technologies was studied in \cite {Zhang2020}. It should be emphasized that the FSO system  experiences  entirely different fading channels (i.e., atmospheric turbulence) compared with the THz.
 
\subsection{Notations and Organization}
We list the main notations  in Table I. This paper is organized as follows: In Section II, we discuss the THz-RF relay system model. In Section III, we analyze the performance of relay assisted system by deriving closed-form expressions for the outage probability, moments of SNR, ergodic rate, and average BER. In Section IV, simulation results of the proposed system are presented. Finally, in Section V, we provide conclusions. 

\section{System Model}\label{sec:system_model}	
	We consider a wireless system where a user terminal lies in a highly shadowed region  to the source, which precludes  RF transmissions. The source can be  a central processing unit of the core network of a cell-free wireless network. To facilitate transmissions, we use an  AP for dual-hop relaying using the DF protocol   between the source and the destination.  We establish the communication of the first hop (i.e., between the source  and the relay) by the THz transmission while the RF is used in  the second hop (i.e.,  between the AP and the user). We consider the generalized $\alpha-\mu$ short-term fading for both the THz and RF links. However, we assume    i.ni.d.  fading    by considering $\alpha_1$, $\mu_1$ and   $\alpha_2$, $\mu_2$ to model the THz and RF links, respectively. We also consider  pointing errors as a channel impairment for the THz link in addition to the short term fading and path loss.

	In the first hop,  the received signal $y_1$ at the relay is expressed as 	
\begin{equation}
	y_1 = h_{l,1}h_{p,1}h_{f,1}s + w_1
\end{equation}		
	where $s$ is the transmitted   signal  in the THz band and $w_1$  is  the additive noise with variance $\sigma_{w1}^2$. The terms $h_{f,1}$ and $h_{p,1}$ model the channel coefficient due to short term fading and pointing errors, respectively.   The deterministic path gain $h_{l,1}$ is dependent on antenna gains, frequency, and  molecular absorption coefficient as given in \cite{Boulogeorgos_Analytical}:	
\begin{equation}
h_l = \frac{c\sqrt{G_{t}G_{r}}}{4\pi f d} \exp(-\frac{1}{2}k(f,T,\psi,p)d)
\end{equation}
	where $c$, $f$, and $d$ respectively denote the speed of light,	the transmission frequency and distance. $G_{t}$, and $G_{r}$ are the antenna gains of the  transmitting antenna and receiving antenna, respectively. The term  $k(f,T,\psi,p)$ is the molecular absorption coefficient  depends on the temperature $T$, relative humidity $\psi$ and atmospheric pressure $p$	
\begin{eqnarray}
	k(f,T,\psi,p) = \frac{q_1v(q_2v+q_3)}{(q_4v+q_5)^2 + (\frac{f}{100c}-p_1)^2} \nonumber \\ +  \frac{q_6v(q_7v+q_8)}{(q_9v+q_{10})^2 + (\frac{f}{100c}-p_2)^2} \nonumber \\ +  c_1f^3 + c_2f^2 + c_3f + c_4
\end{eqnarray}
	where $v = \frac{\psi}{100} \frac{p_w(T,p)}{p}$. The term $p_w(T,p)$ represents the saturated water vapor partial pressure at temperature $T$, and can be evaluated based on Buck’s equation. The values of the other parameters are given in Table II \cite{Boulogeorgos_performance_2018}.
		 
	The probability distribution function (PDF) of pointing errors fading $h_{p,1}$ is given as \cite{Farid2007}:
\begin{equation}
	\begin{aligned}
	f_{h_{p,1}}(h_p) &= \frac{\phi^2}{S_{0}^{\phi^2}}h_{p}^{\phi^{2}-1},0 \leq h_p \leq S_0,
	\end{aligned}
	\label{eq:pdf_hp}
\end{equation}
	where $S_0=\mbox{erf}(\upsilon)^2$ with $\upsilon=\sqrt{\pi/2}\ (r_1/\omega_z)$ and $\omega_z$ is the beam-width, $\phi = {\frac{\omega_{z_{\rm eq}}}{2 \sigma_{s}}}$ with  $\omega_{z_{\rm eq}}$ as the equivalent beam-width at the receiver, which is given as $\omega_{z_{\rm eq}}^2 = {\omega^2_z} \sqrt{\pi} \mbox{erf}(\upsilon)/(2\upsilon\exp(-\upsilon^2)) $, and $\sigma_{s}$ is the variance of pointing errors displacement characterized by the horizontal sway and elevation \cite{Farid2007}.
	
The short term fading 	$|h_f|$ with the $\alpha-\mu$ distribution for  the THz link is given as	 
\begin{equation} \label{eqn:pdf_hf_thz}
	f_{|h_{f,1}|}(x) = \frac{\alpha \mu_1^{\mu_1}}{ \Omega^{\alpha_1\mu_1}_f\Gamma (\mu_1)} x^{\alpha_1\mu_1-1} \exp \bigg(-\mu_1 \frac {x^{\alpha_1}}{\Omega^{\alpha_1\mu_1}_f}\bigg)
\end{equation}
where $\Omega$ is the $\alpha$-root mean value of the fading channel envelope.
	 
Using \eqref{eq:pdf_hp} and \eqref{eqn:pdf_hf_thz},  the PDF of the combined short term fading and pointing errors $|h_{pf}|= |h_p h_f|$ is given as \cite{Boulogeorgos_Analytical}:
\begin{eqnarray} \label{eqn:pdf_hfp_thz}
	f_{|h_{fp}|}(x) = \phi S_0^{-\phi} \frac{\mu_1^{\frac{\phi}{\alpha_1}}  x^{\phi-1} }{\Omega^{\alpha_1} \Gamma (\mu_1)} \Gamma \bigg(\frac{\alpha_1 \mu_1 - \phi}{\alpha_1}, \mu_1 \frac{S_0^{-\alpha_1}}{\Omega^{\alpha_1}} x^{\alpha_1} \bigg)
\end{eqnarray}	
	
\begin{figure}[tp]	
	\centering
	\includegraphics[width=\columnwidth]{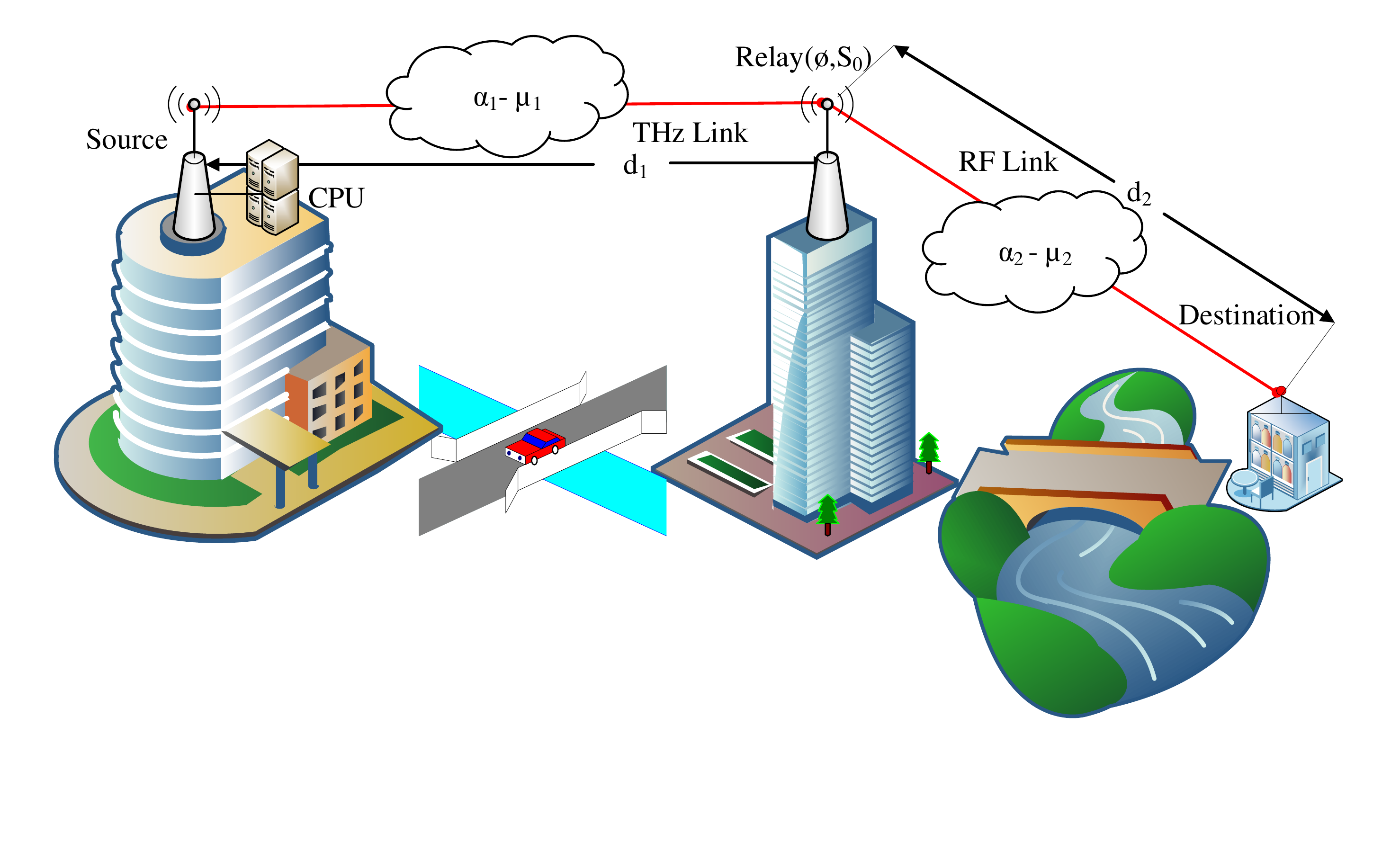}
	\vspace{-15mm}
	\caption{Relay assisted THz-RF wireless link.}
	\label{system_model}	
\end{figure}
In the second hop,  assuming that the signal received through the direct link is negligible, the received signal $y_2$ at the destination is expressed as 	
	\begin{equation}
	y_2 = h_{l,2}h_{f,2}\hat{s} + w_2
	\end{equation}	
where $\hat{s}$ is the decoded symbol at the relay, $ h_{l,2} $ is the channel path gain of RF link,  $w_2$ is the additive noise with variance $\sigma_{w2}^2$ and  $h_{f,2}$ is the  short-term  fading with $\alpha$-$\mu$ distribution for the RF link, which is	 
\begin{equation} \label{eqn:pdf_hf_rf}
	 f_{|h_{f,2}|}(x) = \frac{\alpha \mu_2^{\mu_2}}{ \Omega^{\alpha_2\mu_2}_f\Gamma (\mu_1)} x^{\alpha_2\mu_2-1} \exp \bigg(-\mu_2 \frac {x^{\alpha_2}}{\Omega^{\alpha_2\mu_2}_f}\bigg)
\end{equation}	 

In what follows, we use the distribution functions of channel fading and pointing errors to analyze the performance of relay-assisted THz-RF system.		
\begin{table}[tp] 
		\caption{List of Symbols}
		\label{table1}
		\vspace{-.4cm}
	 \begin{center}
	 	\begin{tabular}{ | c | c|} 
	 		\hline
	 		\textbf{Symbol} & \textbf{Parameter} \\ 
	 		\hline
	 		$(.)_1$ & Notation for the first link \\ 
	 		\hline
	 		$(.)_2$ & Notation for the second link \\ 
	 		\hline
	 		$\phi, S_0$ & Pointing error parameter \\ 
	 		\hline
	 		$\alpha, \mu, \Omega$& Fading parameters \\ 
	 		\hline
	 		$y$ & Received signal \\ 
	 		\hline
	 		$h$ & Channel coefficient \\ 
	 		\hline
	 		$s, \hat{s}$ & Transmitted signal \\ 
	 		\hline
	 		$w$ & Additive Noise \\ 
	 		\hline
	 		$h_f$ & Short term fading for RF link \\ 
	 		\hline
	 		$h_{p}$ & Pointing error for THz link \\ 
	 		\hline
	 		$f(x)$ & Probability distribution function \\ 
	 		\hline
	 		$F(x)$ & Cumulative distribution function \\ 
	 		\hline
	 		$\gamma$ & Signal to noise ratio \\ 
	 		\hline
	 		$\gamma^0$ & Signal to noise ratio without fading \\ 
	 		\hline
	 		$\eta$ & Ergodic Capacity \\ 
	 		\hline
	 		$P_e$ & Probability of error \\ 
	 		\hline
	 		$_2\tilde{F}_1$ & Regularized Hypergeometric function \\ 
	 		\hline
	 		$G_{p,q}^{m,n}(.|.)$ & Meijer's G function \\ 
	 		\hline
	 		$\Gamma(a)$ & $\int_{0}^{\infty} t^{a-1}e^{-t}dt$ \\ 
	 		\hline
	 		$\Gamma(a,t)$ & $\int_{t}^{\infty} s^{a-1}e^{-s}ds$ \\ 
	 		\hline
	 		$\gamma(a,t)$ & $\int_{0}^{t} s^{a-1}e^{-s}ds$ \\ 
	 		\hline
	 		$Q(.)$ & $\frac{1}{\sqrt{2\pi}} \int_{\gamma}^{\infty} e^{\frac{-u^2}{2}} du$ \\ 
	 		\hline
	 		$\Delta(k,a)$ & $\frac{a}{k}$, $\frac{a+1}{k}$, ..., $\frac{a+k-1}{k}$  \\ 
	 		\hline
	 	\end{tabular}
	 \end{center}
\end{table}
 	 
\section{Performance Analysis}
In this section, we analyze the performance of the relay-assisted system. First, we derive a closed-form expression on the CDF of THz link over i.ni.d $\alpha$-$\mu$ fading channel with pointing error. Next, we use the derived PDFs and CDFs to analyze the outage probability, moments of SNR, ergodic capacity, and average BER performance of the THz-RF relay system. 

To this end, we denote  instantaneous SNR of  the THz link as 	$ \gamma_{1}= \gamma_{1}^0|h_{fp}|^2$ and the instantaneous SNR of RF as $\gamma_{2}=\gamma_{2}^0|h_{f}|^2$,  where $\gamma_{1}^0= \frac{P_t |h_{1}|^2}{\sigma_{w_1}^2}$ and  $\gamma_{2}^0= \frac{P_r |h_{2}|^2}{\sigma_{w_2}^2}$ are the SNR terms  without fading for the THz and RF links, respectively. Using  \eqref{eqn:pdf_hfp_thz}, we can represent the PDF of the THz link in terms  of SNR \cite{papoulis_2002}

\begin{eqnarray}
f_1(\gamma) = \frac{A_1}{2\sqrt{\gamma \gamma_1^0}} \Big(\sqrt{{\gamma}/{\gamma_1^0}}\Big)^{\phi-1} \Gamma \Big(B_1, C_1 \Big(\sqrt{{\gamma}/{\gamma_1^0}}\Big)^{\alpha_1} \Big)
\label{eqn:pdf_thz}
\end{eqnarray}
where $A_1$ = $\phi S_0^{-\phi} \frac{\mu_1^{\frac{\phi}{\alpha_1}}}{\Omega^\alpha_1 \Gamma (\mu_1)} $, $B_1$ = $ \frac{\alpha_1 \mu_1 - \phi}{\alpha_1} $, and $C_1$ = $\frac{\mu_1}{\Omega^\alpha_1} S_0^{-\alpha_1}$.

Similarly, we can express \eqref{eqn:pdf_hf_rf} in terms of  SNR for the RF link as
\begin{eqnarray}
f_2(\gamma) \hspace{-1mm}= \frac{A_2 \alpha_2}{2 \Gamma(\mu_2)\sqrt{\gamma\gamma_2^0}} \Big(\sqrt{{\gamma}/{\gamma_2^0}}\Big)^{(\alpha_2\mu_2-1)} \large{e}^{-B_2 \Big(\sqrt{{\gamma}/{\gamma_2^0}}\Big)^{\alpha_2}}
\label{eqn:pdf_rf}
\end{eqnarray}
where $A_2 = \frac{ \mu_2^{\mu_2}}{\Omega^{\alpha_2\mu_2}}$ and $B_2 = \frac{\mu_2}{\Omega^{\alpha_2}}$.
 
Finally, the CDF of the SNR for the RF link:
\begin{eqnarray}
F_2(\gamma) =  1-\bigg(\frac{\Gamma\big(\mu_2, B_2 \big(\sqrt{{\gamma}/{\gamma_2^0}}\big)^{\alpha_2}\big)}{\Gamma (\mu_2)}\bigg)
\label{eqn:cdf_rf}  
\end{eqnarray}

\subsection{Cumulative Distribution Function of THz Link}
In the following, we derive a closed-form expression of the CDF for  the THz Link.
\begin{my_theorem} \label{th:cdf}
	If $\phi$ and $S_0$ be the parameters of  pointing errors, and  $\alpha_1$ and $\mu_1$ are the fading parameters, then the CDF of the THz link  is given by 
	\begin{flalign}	\label{eqn:cdf_thz}
F_1 (\gamma)=  \frac{A_1  C_1^{-\frac{\phi}{\alpha_1}}}{\phi} &\bigg[ \gamma\bigg(\mu_1,C_1\Big(\sqrt{{\gamma}/{\gamma_1^0}}\Big)^{\alpha_1}\bigg) \hspace{-1mm}+ C_1^{\frac{\phi}{\alpha_1}}\Big(\sqrt{{\gamma}/{\gamma_1^0}}\Big)^{\hspace{-1mm}\phi}\nonumber \\  &\times \Gamma\Big(B_1,C_1\Big(\sqrt{{\gamma}/{\gamma_1^0}}\Big)^{\alpha_1}\Big) \bigg]
\end{flalign}
	
\end{my_theorem}
\begin{IEEEproof} Using  \eqref{eqn:pdf_thz} and substituting $\big(\sqrt{{\gamma}/{\gamma_1^0}}\big)^{\alpha_1} = t$ ,  the CDF of SNR for the THz link is  given by
 \begin{equation}
 F_1 (\gamma) = \frac{A_1}{\alpha_1} \int_{0}^{\big(\hspace{-1mm}\sqrt{{\gamma}/{\gamma_1^0}}\big)^{\alpha_1}} \hspace{-3mm}t^{(\frac{\phi}{\alpha_1}-1)} \Gamma (B_1,C_1 t) dt 
 \label{eq:simp}
\end{equation}

To solve the above integral, we use the following identity:	
\begin{eqnarray}
\int x^{b-1} \Gamma(s, x) \mathrm{d} x= -\frac{1}{b}\big(x^b\Gamma(s,x)+\Gamma(s+b,x)\big)
\label{eq:gamma_inc_identity}
\end{eqnarray} 
Applying the limits of  \eqref{eq:simp} in the indefinite integral of \eqref{eq:gamma_inc_identity}, we get \eqref{eqn:cdf_thz}.
\end{IEEEproof}

It can be seen that the CDF in \eqref{eqn:cdf_thz} consists of gamma and incomplete gamma functions and  is applicable for continuous values of $\mu_1$, in contrast to \cite{Boulogeorgos_Error}, which is applicable only for the integer values of $\mu_1$. Further, the derived CDF might facilitate performance analysis in closed-forms.

Finally, we provide the distribution functions of the relay-assisted system. Since $\gamma_{1}$ and 
$\gamma_{2}$ are independent, the expression of end-to-end SNR with DF relaying is given as:
\begin{eqnarray}
	\gamma = \min(\gamma_{1},\gamma_{2})
\end{eqnarray}
Thus, the CDF and PDF of 	$\gamma$  can be given as \cite{papoulis_2002}
\begin{eqnarray} \label{cdf_relay}
	F(\gamma) = F_1(\gamma)+F_2(\gamma)-F_1(\gamma)F_2(\gamma)
\end{eqnarray}
\begin{eqnarray}\label{pdf_relay}
	f(\gamma) = f_1(\gamma)+f_2(\gamma)-f_1(\gamma)F_2(\gamma)-F_1(\gamma)f_2(\gamma)
\end{eqnarray}
where $f_1(\gamma)$, $f_2(\gamma)$ are the PDF of the THz and RF link, respectively. Similarly, $F_1(\gamma)$ and $ F_2(\gamma) $ are the CDF of the THz and RF link, respectively.

\subsection{Outage Probability}
Outage probability is one of the  important performance metrics to analyze the effect of  fading channels. The outage probability $P_{\rm out}$ is defined as the probability of failing to reach an SNR threshold value 
$\gamma_{th}$, i.e., $ P_{\rm out} = P(\gamma <\gamma_{th}) $. 

\begin{my_corollary}
		\begin{enumerate}[label=(\alph*)]
			\item An exact expression for the outage probability is $P_{\rm out}= F(\gamma_{\rm th})$, where $F(\gamma)$ is given in \eqref{cdf_relay} and $\gamma_{\rm th}$ is the threshold value of SNR.
		\item Asymptotically, 	 for a low SNR regime, an expression of the outage probability is given in \eqref{eq:diversity_eqn_low}.
		\item Asymptotically, for a high SNR regime, an expression of the outage probability is given in \eqref{eq:diversity_eqn_high}.
		\item The diversity order $M$ of the THz-RF relay-assisted system is given by		
		\begin{eqnarray}\label{diversity order}
		M = \min \bigg\{\frac{\alpha_1\mu_1}{2}, \frac{\alpha_2\mu_2}{2}, \frac{\phi}{2} \bigg\}
		\end{eqnarray}
			
			\end{enumerate}
\end{my_corollary}
\begin{figure*}
	\begin{flalign}	\label{eq:diversity_eqn_low}
	&P_{\rm out}\approx 1- \frac{1}{\Gamma(\mu_2)}e^{-B_2 \big({\gamma}/{\gamma_2^{(0)}}\big)^{\frac{\alpha_2}{2}}} {B_2^{\mu_2-1} \big({\gamma}/{\gamma_2^{(0)}}\big)^{\frac{\alpha_2(\mu_2-1)}{2}}} +\frac{A_1  C_1^{-\frac{\phi}{\alpha_1}}}{\phi} \frac{1}{\Gamma(\mu_2)}e^{-B_2 \big({\gamma}/{\gamma_2^{(0)}}\big)^{\alpha_2/2}} {B_2^{\mu_2-1} \big({\gamma}/{\gamma_2^{(0)}}\big)^{\frac{\alpha_2(\mu_2-1)}{2}}} \nonumber\\
	&\bigg[\Big(\Gamma(\mu_1)+ C_1^{\phi/\alpha_1}\Big({\gamma}/{\gamma_1^0}\Big)^{\phi/2}\Big) e^{-C_1 \big({\gamma}/{\gamma_1^{(0)}}\big)^{\alpha_1/2}} {C_1 \big({\gamma}/{\gamma_1^{(0)}}\big)^{\alpha_1/2}\big)} ^{k-1}  - e^{-C_1 \big({\gamma}/{\gamma_1^{(0)}}\big)^{\alpha_1/2}} \Gamma(\mu_1) {C_1^{\mu_1-1} \big({\gamma}/{\gamma_1^{(0)}}\big)^{\frac{\alpha_1(\mu_1-1)}{2}}} \bigg]
	\end{flalign}

	\begin{flalign} 
	{P_{out}}\approx   \bigg(\frac{1}{\mu_1} - \frac{S_0^{\alpha_1\mu_1}\mu_1^{\mu_1}}{\Gamma (\mu_1)B_1 }\bigg) \bigg(\frac{\gamma}{\gamma_1^0}\bigg)^{\hspace{-2mm}\frac{\alpha_1\mu_1}{2}} \hspace{-4mm}+ {{ S_0^{-\phi} \mu_1^{\frac{\phi}{\alpha_1}} \frac{\Gamma(B_1)}{ \Gamma (\mu_1)} } {} \bigg(\frac{\gamma}{\gamma_1^0}\bigg)}^{\frac{\phi}{2}}   +\frac{1}{\Gamma(\mu_2)} \bigg(\frac{\gamma}{\gamma_2^0}\bigg)^{\frac{\alpha_2\mu_2}{2}}
	\label{eq:diversity_eqn_high}
	\end{flalign}
	\hrule
\end{figure*}
\begin{IEEEproof}
	Proof for part (a) is trivial. For part (b), we use  the asymptotic expression of $\lim _{x \to \infty}\Gamma(a,x) = e^{-x}x^{a-1}$ \cite{Jameson2016} in \eqref{cdf_relay} containing  $F_1(\gamma)$ (see \eqref{eqn:cdf_rf}) and $F_2(\gamma)$  (see \eqref{eqn:cdf_thz}). For part (c), we use $F_1{(\gamma_{\rm th})}+F_2{(\gamma_{\rm th})} >> F_1{(\gamma_{\rm th})}F_2{(\gamma_{\rm th})} $ and  $\lim _{x \to 0}\gamma(s,x) = x^s/s$ \cite{Jameson2016}  in \eqref{cdf_relay},  to get \eqref{eq:diversity_eqn_high}. 
For part (d),  the diversity order can be obtained using the exponent of SNR (i.e., dominant terms of the outage probability at high SNR)  in \eqref{eq:diversity_eqn_high}. 
	\end{IEEEproof}

The expression of diversity order in \eqref{diversity order} shows that the effect of pointing errors can be mitigated if the normalized beam width is adjusted sufficiently to get $\phi> \min \{\alpha_1 \mu_1, \alpha_2 \mu_2\}$.

\subsection{Statistical Analysis of  SNR}
In this subsection, we derive the $n$th moment of SNR, which can  be used to analyze other statistical parameters such as average SNR and amount of fading (AoF).  The AoF is a key performance parameter to analyze the severity of channel fading.  

Using \eqref{pdf_relay}, the $n$-th moment of  SNR for the relay assisted system under the considered fading channel is given as
\begin{equation} \label{eq:total_pdf}
 \bar{\gamma}^{(n)} = \int_{0}^{\infty}  \gamma^n f(\gamma)d \gamma
\end{equation} 
In the following theorem, we derive the $n$-th moment of SNR considering the generalized i.ni.d. fading model.

\begin{figure*}	[tp]
	\begin{flalign} \label{eq:gamma_12_asy}
	\bar{\gamma}_{12}^{(n)} = \bar{\gamma}_{1} +  \frac{A_1 C_1^{-\frac{\phi+2n}{\alpha_1}} \hspace{-1mm}{\gamma_{1}^0}^n \alpha_1^{(\mu_2-\frac{1}{2})} \alpha_2^{(B_1+\frac{\phi+2n}{\alpha_1}-\frac{3}{2})}}{\Gamma(\mu_2)\alpha_1 (2\pi)^{(\frac{\alpha_1+\alpha_2-2}{2})}} G_{\alpha_1+2\alpha_2, 2\alpha_1+\alpha_2}^{2\alpha_1,2\alpha_2} \Bigg(\hspace{-1mm}\begin{matrix} \Delta\big(\alpha_2, 1\hspace{-0.5mm}-\hspace{-0.5mm}(\frac{\phi+2n}{\alpha_1}\hspace{-0.5mm}-\hspace{-0.5mm}B_1)\big),\Delta\big(\alpha_2, 1\hspace{-0.5mm}-\hspace{-0.5mm}(\frac{\phi+2n}{\alpha_1})\big), \Delta(\alpha_1,1) \\ \Delta(\alpha_1, \mu_2),\Delta(\alpha_1,0),\Delta\big(\alpha_2, -(\frac{\phi+2n}{\alpha_1})\big) \end{matrix} \Bigg| \frac {B_2'^{\alpha_1} \alpha_1^ {-\alpha_1}} {C_1^{\alpha_2} \alpha_2^{-\alpha_2}}\hspace{-1mm}\Bigg) 
	\end{flalign}
\end{figure*}
\begin{figure*}	[tp]
	\begin{flalign}  \label{eq:gamma_21_asy}
	\bar{\gamma}_{21}^{(n)} &\hspace{-1mm}=\hspace{-1mm} \frac{A_1 C_1^{\frac{-\phi}{\alpha_1}} \Gamma(\mu_1)}{\phi}\bar{\gamma}_{2} +\hspace{-1mm}  \frac{A_1A_2 C_1^{\frac{-\phi}{\alpha_1}} C_1' {\gamma_{2}^0}^n \hspace{-0.5mm}\alpha_1^{(\mu_2-\frac{\phi+2n}{\alpha_2}-\frac{1}{2})} \hspace{-0.5mm}\alpha_2^{(B_1-\frac{1}{2})} \hspace{-0.5mm}B_2^{-(\mu_2+\frac{\phi+2n}{\alpha_2})}}{\Gamma(\mu_2)\phi (2\pi)^{(\frac{\alpha_1+\alpha_2-2}{2})}} G_{\alpha_1+\alpha_2, 2\alpha_2}^{2\alpha_2,\alpha_1}\Bigg(\hspace{-1mm}\begin{matrix} \Delta\big(\alpha_1, \alpha_1\hspace{-1mm}-\hspace{-1mm}(\mu_2\hspace{-1mm}+\hspace{-1mm}\frac{\phi+2n}{\alpha_2})\big), \Delta(\alpha_2,1) \\ \Delta(\alpha_2, B_1),\Delta(\alpha_2,0) \end{matrix} \Bigg| \frac {C_1'^{\alpha_2} \alpha_1^ {\alpha_1}} {B_2^{\alpha_1} \alpha_2^{\alpha_2}}\hspace{-1mm}\Bigg) \nonumber\\ &+ \frac{A_1A_2 C_1^{\frac{-\phi}{\alpha_1}} {\gamma_{2}^0}^n \alpha_1^{(\mu_2+\frac{2n}{\alpha_2}-\frac{1}{2})} \alpha_2^{(\mu_2-1)} B_2^{-(\mu_2+\frac{2n}{\alpha_2})}}{\Gamma(\mu_2)\phi (2\pi)^{(\frac{\alpha_1+\alpha_2-2}{2})}}  G_{\alpha_1+\alpha_2, 2\alpha_2}^{2\alpha_2,\alpha_1}\Bigg(\begin{matrix} \Delta\big(\alpha_1, \alpha_1-(\mu_2+\frac{2n}{\alpha_2})\big), \Delta(\alpha_2,1) \\ \Delta(\alpha_2, \mu_1),\Delta(\alpha_2,0) \end{matrix} \Bigg| \frac {C_1'^{\alpha_2} \alpha_1^ {\alpha_1}} {B_2^{\alpha_1} \alpha_2^{\alpha_2}}\Bigg)	
	\end{flalign} 
	\hrule
\end{figure*}			
 \begin{my_theorem}  \label{th:snr_asym}
  Let $\phi$ and $S_0$ be the parameters of pointing errors of the THz link. If  $\alpha_1$,  $\mu_1$ and  $\alpha_2$,  $\mu_2$ are  the fading parameters of THz and RF links, respectively,  then the $n$-th moment of  SNR of the relay assisted THz-RF link is given as:
 \begin{eqnarray}  \label{eqn:total_snr_asy}
 	\bar{\gamma}^{(n)} =\bar{\gamma}_{1}^{(n)}+\bar{\gamma}_2^{(n)}-\bar{\gamma}_{12}^{(n)}-\bar{\gamma}_{21}^{(n)}
 \end{eqnarray} 		
 	where 
 \begin{equation} 
 	\bar{\gamma}_1^{(n)}\hspace{-1mm} =\hspace{-1mm}  \frac{A_1C_1^{\hspace{-0.5mm} -(\hspace{-0.5mm} \frac{\phi+2n}{\alpha_1}\hspace{-0.5mm} )} \hspace{-1mm}{\gamma_1^0}^n \Gamma(\hspace{-0.5mm} \frac{\alpha_1 B_1+\phi+2n}{\alpha_1}\hspace{-0.5mm} )}{2n+\phi}, 
 	 \bar{\gamma}_2^{(n)} \hspace{-1mm}=\hspace{-1mm} \frac{B_2^{\hspace{-0.5mm} \hspace{-0.5mm} -\frac{2n}{\alpha_2}} {\gamma_2^0}^n \Gamma(\hspace{-0.5mm} \frac{2n}{\alpha_2}\hspace{-1mm}+\hspace{-1mm}\mu_2\hspace{-0.5mm} )}{\Gamma(\mu_2)},
 \label{eq:gamma_1_and_2_asy}
 \end{equation}
$\bar{\gamma}_{12}^{(n)}$, and $\bar{\gamma}_{21}^{(n)}$ are given in \eqref{eq:gamma_12_asy} and \eqref{eq:gamma_21_asy}, respectively.
 \end{my_theorem}
 \begin{IEEEproof}
 	The proof is presented in Appendix A.
 \end{IEEEproof}
It should be noted that $\bar{\gamma}_1^{(n)}$ and $\bar{\gamma}_2^{(n)}$ are the $n$-th moment of SNR for the THz and RF links, respectively. In the following Lemma 1, we describe  a special case of Theorem 2 by considering the  i.i.d model where short-term fading between source to relay and relay to destinations is identical.
\begin{my_lemma} 
	Let $\phi$ and $S_0$ be the parameters of  pointing errors, and  $\alpha$, $\mu$ be the fading parameters, then the $n$-th moment of  SNR of the relay assisted THz-RF link is given as
	\begin{eqnarray}  \label{eqn:total_snr}
	\bar{\gamma}^{(n)} =\bar{\gamma}_{1}^{(n)}+\bar{\gamma}_2^{(n)}-\bar{\gamma}_{12}^{(n)}-\bar{\gamma}_{21}^{(n)}
	\end{eqnarray} 		
	where 
	\begin{equation} 
		\bar{\gamma}_1^{(n)} \hspace{-1mm}= \hspace{-0.5mm}  \frac{A_1C_1^{-(\frac{\phi+2n}{\alpha})} {\gamma_1^0}^n \Gamma(\frac{\alpha B_1+\phi+2n}{\alpha})}{2n+\phi},  \hspace{-0.5mm}  \bar{\gamma}_2^{(n)}  \hspace{-1mm}=\hspace{-0.5mm}  \frac{B_2^{-\frac{2n}{\alpha}} \gamma_2^0 \Gamma(\frac{2n}{\alpha}\hspace{-1mm}+\hspace{-1mm}\mu)}{\Gamma(\mu)}
		\label{eq:gamma_1_and_2}
	\end{equation}	
	\begin{flalign} \label{eq:gamma_12}
		&\bar{\gamma}_{12}^{(n)}  = \bar{\gamma}_1^{(n)} - \frac{1}{(\phi+2n)\Gamma(\mu)}\bigg[A_1 B_2'^{-\frac{\phi+2n}{\alpha} }{\gamma_{1}^0}^n \bigg(\hspace{-1mm}\Gamma(B_1) \Gamma\hspace{-1mm}\left(\frac{\alpha\mu+\phi+2n}{\alpha}\right) \nonumber \\& +B_2'^{-B_1} C_1^{B_1} \Gamma\left(\frac{\alpha(B_1+\mu)+\phi+2n}{\alpha}\right) \nonumber \\& \times \bigg(\frac{\alpha ~ _2\tilde{F}_1\big[\frac{\alpha B_1+\phi+2n}{\alpha}, \frac{\alpha(B_1+\mu)+\phi+2n}{\alpha}, \frac{\alpha B_1+\alpha+\phi+2n}{\alpha}, -\frac{C_1}{B_2'} \big]}{\alpha B_1+\phi+2n} \nonumber \\&  - \hspace{-1mm}\Gamma(B_1)~ \hspace{-1mm}_2\tilde{F}_1 \bigg[B_1, \frac{\alpha(B_1+\mu)+\phi+2n}{\alpha}, 1+B_1, -\frac{C_1}{B_2'} \bigg] \bigg) \bigg) \bigg]
	\end{flalign}	
	\begin{flalign}  \label{eq:gamma_21}
		&\bar{\gamma}_{21}^{(n)} =  \frac {A_1 C_1^{\frac{-\phi}{\alpha}} \Gamma(\mu)  \bar{\gamma}_{2}^{(n)}}{\phi} + \frac{1}{\phi\Gamma(\mu)}\bigg[\alpha A_1 A_2 C_1^{-\frac{\phi}{\alpha}} C_1'^{-\frac{\alpha\mu+\alpha+2n}{\alpha}} {\gamma_2^0}^n \nonumber \\& \times \bigg(\hspace{-2mm}-\frac{\Gamma(1\hspace{-1mm}+ \frac{2n}{\alpha}\hspace{-1mm}+2\mu) ~ _2\tilde{F}_1\big[1+ \frac{2n}{\alpha}+\mu, 1+ \frac{2n}{\alpha}+2\mu, 2+ \frac{2n}{\alpha}+\mu, -\frac{B_2}{C_1'} \big]}{\alpha \mu +\alpha	+2n} \nonumber \\ & + (\alpha \mu +\alpha + \phi +2n)^{-1}\Gamma\Big(\frac{\alpha(\mu+B_1+1)+\phi+2n}{\alpha}\Big)  \nonumber\\&  \times {_2\tilde{F}_1} \Big[\frac{\alpha\mu\hspace{-1mm}+\hspace{-1mm}\alpha\hspace{-1mm}+\hspace{-1mm}\phi\hspace{-1mm}+\hspace{-1mm}2n}{\alpha}, \frac{\alpha(\mu\hspace{-1mm}+\hspace{-1mm}B_1\hspace{-1mm}+\hspace{-1mm}1)\hspace{-1mm}+\hspace{-1mm}\phi\hspace{-1mm}+\hspace{-1mm}2n}{\alpha}, \frac{\alpha(\mu\hspace{-1mm}+\hspace{-1mm}2)\hspace{-1mm}+\hspace{-1mm}\phi\hspace{-1mm}+\hspace{-1mm}2n}{\alpha}, -\frac{B_2}{C_1'} \Big] \bigg) \bigg] 	
	\end{flalign} 				
\end{my_lemma}
\begin{IEEEproof}
	The proof is presented in Appendix B.
\end{IEEEproof}

The derived expressions of the $n$-th moment of SNR in Theorem 2 and Lemma 1 can be used to find 
the average SNR of the relay assisted link  by substituting  $n=1$ in \eqref{eqn:total_snr_asy} and \eqref{eqn:total_snr}. Further, a closed-form expression for the $2^{\rm nd}$ order AoF is expressed as
\begin{eqnarray}
	\rm {AoF} = \frac{\bar{\gamma}^{(2)}}{(\bar{\gamma}^{(1)})^2}-1
\end{eqnarray} 

Although, the derived expressions in \eqref{eqn:total_snr_asy} and \eqref{eqn:total_snr} involves standard mathematical functions, it  is desirable to simplify further  the analysis under  practically relevant scenarios in order to provide insights on the system behavior.
Thus, we present  much simplified expressions of the average SNR  by considering some specific fading  conditions and pointing error parameters in the following Corollary 2.
\begin{my_corollary}
	\begin{enumerate}[label=(\alph*)]
		\item  Considering the THz-link as the Nakagami-$m$ fading ($\alpha_1=2$, $\mu_1=2$) with pointing error parameter   $\phi=2$, and the  Rayleigh fading ($\alpha_1=2$ $\mu_1=1$) for the RF-link,  the average SNR of the relay assisted link is given as 
				
		\begin{flalign} \label{eq:special_snr1}
		\bar{\gamma} = & \gamma_{1}^0 + 2\gamma_2^0 - 2S_0^{-2} \gamma_1^0\bigg(\gamma_1^0 - \frac{1}{(\sqrt{{\gamma_1^0}/{\gamma_2^0}}+2S_0^{-2})^2} \bigg)\nonumber \\ &- \gamma_2^0 \bigg(\frac{2(1+2S_0^{-2})^4+6-(1+2S_0^{-2})^2}{(1+2S_0^{-2})^4} \bigg)
		\end{flalign} 		
		\item Considering the THz-link ($\alpha_1$$\to$$\infty$, $\mu_1=1$) as a linear Weibull fading with  $\phi=2$, and the  Rayleigh fading ($\alpha_2=2$ $\mu_2=1$) for the RF-link,  the average SNR of the relay assisted link:	
		\begin{flalign}\label{eq:special_snr2}
		\bar{\gamma} = \gamma_{2}^0 - \frac{S_0^{-\phi}\gamma_{2}^0}{\phi}
		\end{flalign} 		
	\end{enumerate}
\end{my_corollary}

\begin{IEEEproof}
Using $n$=$1$ and $f(\gamma)$ from \eqref{pdf_relay} in  \eqref{eq:total_pdf} with special cases of $\alpha$ and $\mu$ enumerated in (a) and (b), and applying the standard procedure, we can get \eqref{eq:special_snr1} and \eqref{eq:special_snr2}.
\end{IEEEproof}
The special case in \eqref{eq:special_snr1} considers popular fading models, which  significantly simplifies the analysis comparing with Theorem \ref{th:snr_asym}.  Further, the scenario in (b)  is chosen to provide insights on the performance asymptotically. As the parameter $\alpha_1$ $\to$ $\infty$ (highly linear behavior i.e., good channel conditions), the performance should be determined by the RF link with finite $\alpha_2$, which is verified using \eqref{eq:special_snr2}.

\subsection{Ergodic Capacity} 
Ergodic capacity is another important performance measure which denotes maximum transmission rate with infinitesimal probability of error:
 \begin{eqnarray} \label{eq:total capacity_pdf}
\bar{\eta}= \int_{0}^{\infty}  \log_2(1+\gamma) f(\gamma) d\gamma 
\end{eqnarray}

The ergodic capacity can also be  expressed using the CDF of SNR \cite{Annamalai2010}
 \begin{eqnarray} \label{eq:total capacity_cdf}
\bar{\eta}= \frac{1}{\rm ln(2)} \int_{0}^{\infty} (1+\gamma)^{-1} (1-F(\gamma)) d\gamma 
\end{eqnarray}

We derive a lower bound on the  ergodic capacity $\bar{\eta}_1$ of the direct THz link to compare with the relay-assisted THz-RF link.
\begin{my_proposition}
	 If $\phi$ and $S_0$ are the parameters of  pointing errors, and $\alpha_1$ and $\mu_1$  are the fading parameters, then  a lower bound on the ergodic capacity of the THz link is:
	 
\begin{flalign}\label{eq:eta_1_asy}
	\bar{\eta}_1 &\geq \frac {1}{\alpha_1 \phi^2 \rm log(2)} \Big[A_1 C_1^{-(\frac{\phi}{\alpha_1})} \Gamma(\mu_1)\big(-2(\alpha_1+\phi \rm log(C_1)) \nonumber \\ &+ \alpha_1\phi \rm log(\gamma_0) + 2\phi\psi(0,\mu_1) \big)\Big]
\end{flalign}	
\end{my_proposition}

\begin{IEEEproof}
	Using (\ref{eqn:pdf_thz}) in \eqref{eq:total capacity_pdf}, and applying the inequality $\log_2(1+\gamma)\geq  \log_2\gamma$ with a substitution $\big(\sqrt{{\gamma}/{\gamma_1^{0}}}\big)^{\alpha_1} = t$, we get:	
	\begin{equation} \label{eq:eta_1_int_asy}
	\bar{\eta}_1\geq \frac{A_1}{\alpha_1}\int_{0}^{\infty} {\rm log_2} (\gamma_1^0 t^{\frac{2}{\alpha_1}}) t^{(\frac{\phi}{\alpha_1}-1)}  \Gamma (B_1,C_1 t) dt 
	\end{equation}	
	To find a closed-form expression,  we use integration by parts taking $\Gamma (B_1,C_1 t)$ as the first and $ {\rm log} (\gamma_1^0 t^{\frac{2}{\alpha_1}}) t^{(\frac{\phi}{\alpha_1}-1)}$ as the second term,  and apply the identity [\cite{Gradshteyn} (eq.4.352/1)] to get $\bar{\eta}_1$ of \eqref{eq:eta_1_asy}. 	
\end{IEEEproof}

It should be noted that the authors in \cite{Boulogeorgos_Analytical} have derived an upper bound on the ergodic capacity of the THz link. For the completeness of analysis,  we present a lower bound on the ergodic  capacity $\bar{\eta}_2$ of the RF link  as
	\begin{equation}\label{eq:eta_2_asy}
	\bar{\eta}_2 \geq \frac{-2 \rm log(B_2)+ \alpha_2 \rm log(\gamma_0) + 2\psi(0,\mu_2)}{\alpha_2 \rm log(2)}
	\end{equation}		

\begin{figure*}
	\begin{flalign}\label{eq:eta_asy}
	&{\cal{F}}_{\eta}(\alpha_1, \alpha_2, \mu_1, \mu_2, \phi, S_0) = \frac{A_1 C_1^{-\frac{\phi}{\alpha_1}} 2^{(\mu_2-\frac{1}{2})}}{\Gamma(\mu_2) \rm {ln(2)} (2\pi)^{(\alpha_2-\frac{1}{2})}} G_{2+\alpha_2, \alpha_2}^{4+\alpha_2,\alpha_2}\Bigg(\begin{matrix} \Delta(\alpha_2,0),\Delta(2,1),   \\ \Delta(\alpha_2, 0),\Delta(2, \mu_2),\Delta(2,0) \end{matrix} \Bigg| \frac {(B_2{\gamma_2^0}^{-\frac{\alpha_2}{2}})^2} {2^{-2}}\Bigg) \nonumber \\ -& \frac{A_1 C_1^{-\frac{\phi}{\alpha_1}} 2^{(\mu_2-\frac{1}{2})} \Gamma(\mu_1)}{\phi \Gamma(\mu_2) \rm {ln(2)} (2\pi)^{(\alpha_2-\frac{1}{2})}} G_{2+\alpha_2, \alpha_2}^{4+\alpha_2,\alpha_2}\Bigg(\begin{matrix} \Delta(\alpha_2,0),\Delta(2,1),   \\ \Delta(\alpha_2, 0),\Delta(2, \mu_2),\Delta(2,0) \end{matrix} \Bigg| \frac {(B_2{\gamma_2^0}^{-\frac{\alpha_2}{2}})^2} {2^{-2}}\Bigg) \nonumber \\  -& \sum_{k_2=0}^{\mu_2-1}\frac{2\sqrt{2}A_1  C_1^{-\frac{\phi}{\alpha_1}+1}{\gamma_1^0} ^{\frac{-\phi}{2}} (B_2 {\gamma_{2}^0}^{-\frac{\alpha_2}{2}})^{k_2} (e^{-B_2{\gamma_{2}^0}^{-\frac{\alpha_2}{2}}}) (C_1{\gamma_1^0} ^{\frac{-\alpha_1}{2}})^{B_1-1} }{ {(2\pi)}^{(\epsilon\alpha_1-\frac{1}{2})}\phi \Gamma(\mu_2) k_2!} G_{\epsilon\alpha_1,2+\epsilon\alpha_1 }^{2+\epsilon\alpha_1,\epsilon\alpha_1}\Bigg(\begin{matrix} \Delta(\epsilon\alpha_1,1-\frac{\phi+k_2\alpha_2+\alpha_1(B_1-1)+2}{2})   \\ \Delta(2, 0)\end{matrix} \Bigg| \frac {(C_1{\gamma_1^0}^{-\frac{\alpha_1}{2}})^2} {4}\Bigg) \nonumber \\ +& \sum_{k_1 = 0}^{\mu_1-1} \sum_{k_2 = 0}^{\mu_2-1} \frac{A_1  C_1^{-\frac{\phi}{\alpha_1}} \Gamma(\mu_1) (C_1 {\gamma_{1}^0}^{-\frac{\alpha_1}{2}})^{k_1} (B_2 {\gamma_{2}^0}^{-\frac{\alpha_2}{2}})^{k_2} (e^{-C_1{\gamma_{1}^0}^{-\frac{\alpha_1}{2}}}) (e^{-B_2{\gamma_{2}^0}^{-\frac{\alpha_2}{2}}})  }{(2\pi)^{(\epsilon\alpha_1-\frac{1}{2})}\phi \rm {ln(2)} {k_1}!{k_2}!}  G_{\epsilon\alpha_1,2+\epsilon\alpha_1}^{2+\epsilon\alpha_1,\epsilon\alpha_1}\bigg(\begin{matrix} \Delta(\epsilon\alpha_1,1-{\frac{k_1 \alpha_1 + k_2 \alpha_2 +2}{2}})  \\ \Delta(2,0) \end{matrix} \bigg|\frac{1}{4 }\bigg)
	\end{flalign}
	\hrule
\end{figure*}	

Now, we derive analytical expressions of the ergodic capacity of the relay-assisted system by considering the CDF-based formula in \eqref{eq:total capacity_cdf}. It can be seen from $\eqref{cdf_relay}$ that  the CDF-based approach reduces the number of computations. We assume that the relay requires little time to relay the data.
In the following Theorem 3, we develop a tight approximation on the ergodic capacity for the i.ni.d fading model.
\begin{my_theorem}  Let $\phi$ and $S_0$ be the parameters of  pointing errors of the THz link. If  $\alpha_1$,  $\mu_1$ and  $\alpha_2$,  $\mu_2$ are  the fading parameters of THz  and RF links, respectively,  then an expression for the ergodic capacity of the relay assisted link is given by
	\begin{eqnarray}
	\bar{\eta}\approx {\cal{F}}_{\eta}(\alpha_1, \alpha_2, \mu_1, \mu_2, \phi, S_0)
	\end{eqnarray}	
where ${\cal{F}}_{\eta}(\alpha_1, \alpha_2, \mu_1, \mu_2, \phi, S_0)$ is given in	 \eqref{eq:eta_asy}.
\end{my_theorem}	
\begin{IEEEproof}
	The proof is presented in Appendix C.
\end{IEEEproof}

Note that the approximation is quite close to the exact result since the derivation uses simple approximation in few terms. In the following Lemma 2, we simplify the analysis of ergodic capacity considering  the i.i.d fading model.

\begin{my_lemma} If $\phi$ and $S_0$ are the parameters of  pointing errors, and  $\alpha$, $\mu$ are the fading parameters, then an expression on the ergodic capacity of the relay assisted THz-RF link is: 
	\begin{eqnarray}
	\bar{\eta}= \bar{\eta}_1+\bar{\eta}_2-\bar{\eta}_{12}-\bar{\eta}_{21},
	\end{eqnarray}
	where		
\begin{flalign} \label{eq:eta_1}
	\bar{\eta}_1\geq& \frac{A_1 C_1^{-\frac{\phi}{\alpha}} \Gamma(\mu)\big(\hspace{-1mm}-\hspace{-1mm}2(\alpha\hspace{-1mm}+\phi \rm log(C_1))\hspace{-1mm} + \hspace{-1mm}\alpha\phi \rm log(\gamma_1^0)\hspace{-1mm} +\hspace{-1mm} 2\phi\psi(0,\mu) \big)}{\alpha \phi^2 \rm log(2)}
\end{flalign}
\begin{flalign}\label{eq:eta_2}
	\bar{\eta}_2\geq\frac{-2  \log(B_2)+ \alpha \rm log(\gamma_2^0) + 2\psi(0,\mu)}{\alpha \rm log(2)}
\end{flalign}	
\begin{flalign} \label{eq:eta_12}
	\bar{\eta}_{12}  &\approx\bar{\eta}_1\hspace{-1mm}+\hspace{-2mm}\sum_{k=0}^{\mu-1}\hspace{-1mm} \frac{2A_1 B_2'^{-\frac{\phi}{\alpha}}}{\alpha^2 \rm log(2) k!}  G_{0,0:2:0}^{0,1:4:2}\bigg(\hspace{-1mm}\begin{matrix}	1\hspace{-1mm}-\hspace{-1mm}\frac{\phi}{\alpha}\hspace{-1mm}-K \\ - \end{matrix} \bigg |\begin{matrix} 1 \\ B_1\hspace{-0.5mm},\hspace{-1mm} 0 \end{matrix} \bigg |\begin{matrix} 1,\hspace{-1mm}1 \\ 1,\hspace{-1mm} 0 \end{matrix} \bigg| \hspace{-1mm}\frac{C_1}{B_2'},\hspace{-1mm}\frac{(\gamma_1^0)^{\frac{\alpha}{2}}}{B_2'}\hspace{-1mm}\bigg)
\end{flalign}
\begin{flalign}\label{eq:eta_21}
	\bar{\eta}_{21} &\approx  \frac{\big(A_1 C_1^\frac{\phi}{\alpha} \Gamma(\mu)\big) }{\phi}\bar{\eta}_2  + \frac{2A_1A_2 C_1^{-\frac{\phi}{\alpha}}  C_1'B_2^{-(\mu+\frac{\phi}{\alpha})}}{\Gamma(\mu)\phi{ \log(2)}\alpha} \nonumber \\ & \times G_{0,0:2:0}^{0,1:4:2} \bigg(\begin{matrix}	1-\mu-\frac{\phi}{\alpha} \\- \end{matrix} \bigg |\begin{matrix} 1 \\ B_1, 0 \end{matrix} \bigg |\begin{matrix} 1,1 \\ 1, 0 \end{matrix} \bigg| \frac{C_1'}{B_2},\frac{(\gamma_2^0)^{\frac{\alpha}{2}}}{B_2}\bigg) \nonumber \\ - &\hspace{-1mm}\sum_{k=0}^{\mu-1} \frac{2A_1A_2  C_1^{-\frac{\phi}{\alpha}}C_1'^k (B_2\hspace{-0.5mm}+\hspace{-0.5mm}C_1')^{\mu+k}}{\alpha \phi{\rm log(2)}k!}  G_{3,2}^{1,3} \bigg(\hspace{-1mm}\begin{matrix} 1,\hspace{-0.5mm}1,\hspace{-0.5mm}1\hspace{-1mm}-\hspace{-1mm}\mu \\ 1,0 \end{matrix} \bigg|\hspace{-0.5mm}\frac{(\gamma_2^0)^{\frac{\alpha}{2}}}{B_2+C_1'} \hspace{-1mm}\bigg) 
	\end{flalign}
\end{my_lemma}	
\begin{IEEEproof}
	The proof is presented in Appendix D.
\end{IEEEproof}

It should be noted that $\bar{\eta}_1$ and $\bar{\eta}_2$ are expressions of the ergodic capacity  for the THz and RF links, respectively.
Similar to the average SNR performance, we present simplified analysis on the ergodic capacity for some specific fading channel conditions and pointing error parameters in the following:
\begin{my_corollary}
	\begin{enumerate}
		[label=(\alph*)]
		\item   Considering the THz-link as Nakagami-$m$ fading ($\alpha_1=2$, $\mu_1=2$) with pointing errors parameter $\phi=2$, and the  Rayleigh fading ($\alpha_1=2$ $\mu_1=1$) for the RF-link,  the ergodic capacity of the relay assisted link is given as 	
		\begin{flalign}\label{eq:special_rate1}
		\bar{\eta} = &- \frac{1\hspace{-1mm}-\hspace{-1mm}(\gamma_2^0 + 2S_0^{-2} \gamma_1^0) e^{(\gamma_2^0 + 2S_0^{-2} \gamma_1^0)} \Gamma(0,\gamma_2^0 + 2S_0^{-2} \gamma_1^0)}{\rm{ln}(2)(\gamma_2^0 + 2S_0^{-2} \gamma_1^0)} \nonumber \\ &+ \frac{e^{(\gamma_2^0 + 2S_0^{-2} \gamma_1^0)} \Gamma(0,\gamma_2^0 + 2S_0^{-2} \gamma_1^0)}{\rm{ln}(2)}
		\end{flalign}
		
		\item Considering the THz-link ($\alpha_1$$\to$$\infty$, $\mu_1=1$) as a linear Weibull fading with  $\phi=2$, and the  Rayleigh fading ($\alpha_2=2$ $\mu_2=1$) for the RF-link,  the ergodic capacity of the relay assisted link: 		
		\begin{flalign}\label{eq:special_rate2}
		\bar{\eta} = \frac{e^{\gamma_2^0} \Gamma(0,\gamma_2^0)} {\rm{ln}(2)}
		\end{flalign}
	\end{enumerate}
\end{my_corollary}

\begin{IEEEproof}
	Using $F(\gamma)$ from \eqref{cdf_relay} in \eqref{eq:total capacity_cdf} with special cases of $\alpha$ and $\mu$ enumerated in (a) and (b), and applying the standard procedure, we can get \eqref{eq:special_rate1} and \eqref{eq:special_rate2}.
\end{IEEEproof}

The special case in \eqref{eq:special_rate1} is a simplified expression of the ergodic capacity with  Nakagami-m and Rayleigh fading channels.  Further, the scenario in (b)  is chosen to show the effect of imbalance of  fading channels in both the links:  as the non-linearity of fading channel reduces to $\alpha$ $\to$ $\infty$  (i.e, good channel conditions), the performance is determined by the another link with finite $\alpha$ (see \eqref{eq:special_rate2}).
\begin{figure*}	
	\begin{flalign} \label{eq:pe_3_asy}
&	{\cal{F}}_{P_e}(p,q,\alpha_1, \alpha_2, \mu_1, \mu_2,\phi, S_0,\epsilon)= \frac{A_1  C_1^{-\frac{\phi}{\alpha_1}} \Gamma(\mu_1)2^{(\mu_2+\frac{1}{2})} \alpha_2^{\frac{2p-1}{2}}}{2\phi\Gamma(p)\Gamma(\mu_2)(2\pi)^{(\alpha_2+1)}}  G_{4,\alpha_2}^{2+\alpha_2,4}\bigg(\begin{matrix} \Delta(\alpha_2,\alpha_2-p), \Delta(2,1) \\ \Delta(2,\mu_2), \Delta(2,0) \end{matrix} \bigg|\frac{(B_2{\gamma_{2}^{0}}^{-\frac{\alpha_2}{2}}){\alpha_2}^{\alpha_2}}{4 q^{(\alpha_2)}}\bigg) \nonumber \\  +& \sum_{k_2=0}^{\mu_2-1}\frac{A_1  C_1^{-\frac{\phi}{\alpha_1}+1}{\gamma_1^0} ^{\frac{-\phi}{2}} (B_2 {\gamma_{2}^0}^{-\frac{\alpha_2}{2}})^{K_2} (e^{-B_2{\gamma_{2}^0}^{-\frac{\alpha_2}{2}}}) (C_1{\gamma_1^0} ^{\frac{-\alpha_1}{2}})^{B_1-1} q^p ({\epsilon\alpha_1})^{(\frac{\phi+k_2 \alpha_2+\alpha_1(B_1-1) +2p-2}{2})} q^{-\frac{\phi+k_2 \alpha_2+\alpha_1(B_1-1)+2p }{2}}}{\sqrt2 \phi\Gamma(p) (2\pi)^{(\epsilon\alpha_1)} k_2!} \nonumber \\ \times& G_{1,2}^{2,1}\bigg(\begin{matrix} \Delta({\epsilon\alpha_1},{\epsilon\alpha_1}-{\frac{\phi+ k_2 \alpha_2 + \alpha_1(B_1-1)+2p }{2}})  \\ \Delta(2,0) \end{matrix} \bigg|\frac{(C_1{\gamma_1^0} ^{\frac{-\alpha_1}{2}})^2 (\epsilon\alpha_1)^{\epsilon\alpha_1}}{4 q^{(\epsilon\alpha_1)}} \bigg)  \nonumber \\  -& \sum_{k_1 = 0}^{\mu_1-1} \sum_{k_2 = 0}^{\mu_2-1} \frac{A_1  C_1^{-\frac{\phi}{\alpha_1}} \Gamma(\mu_1) (C_1 {\gamma_{1}^0}^{-\frac{\alpha_1}{2}})^{k_1} (B_2 {\gamma_{2}^0}^{-\frac{\alpha_2}{2}})^{k_2} (e^{-C_1{\gamma_{1}^0}^{-\frac{\alpha_1}{2}}}) (e^{-B_2{\gamma_{2}^0}^{-\frac{\alpha_2}{2}}}) q^p(\epsilon\alpha_1)^{\frac{k_1 \alpha_1 + k_2 \alpha_2+2p-1}{2}} q^{-\frac{k_1 \alpha_1 + k_2 \alpha_2 +2p}{2}}}{\sqrt2 \phi\Gamma(p) (2\pi)^{(\epsilon\alpha_1)} {k_1}!{k_2}!} \nonumber \\ \times& G_{\epsilon\alpha_1,2}^{2,\epsilon\alpha_1}\bigg(\begin{matrix} \Delta(\epsilon\alpha_1,\epsilon\alpha_1-{\frac{k_1 \alpha_1 + k_2 \alpha_2 +2p}{2}})  \\ \Delta(2,0) \end{matrix} \bigg|\frac{{\epsilon\alpha_1}^{\epsilon\alpha_1}}{4 q^{(\epsilon\alpha_1)}}\bigg)+ \frac{1}{2} -  \frac{2^{(\mu_2+\frac{1}{2})} \alpha_2^{\frac{2p-1}{2}}}{2\Gamma(p)\Gamma (\mu_2)(2\pi)^{(\alpha_2+1)}}G_{4,\alpha_2}^{2+\alpha_2,4}\bigg(\begin{matrix} \Delta(\alpha_2,\alpha_2-p), \Delta(2,1) \\ \Delta(2,\mu_2), \Delta(2,0) \end{matrix} \bigg|\frac{(B_2{\gamma_{2}^{0}}^{-\frac{\alpha_2}{2}}){\alpha_2}^{\alpha_2}}{4 q^{(\alpha_2)}}\bigg)
	\end{flalign}
	\hrule
\end{figure*}
\subsection{Average Bit Error Rate}
In this subsection, we derive the average BER for the  relay-assisted THz-RF link. Using the CDF, the average BER can be expressed as \cite{Ansari2011}:
\begin{eqnarray} \label{eq:ber}
\bar{P_e} = \frac{q^p}{2\Gamma(p)}\int_{0}^{\infty} \gamma^{p-1} {e^{{-q \gamma}}} F (\gamma)   d\gamma
\end{eqnarray}
where $ q $ and $ p $ are modulation-dependent parameters. Specifically, $p=1$ and $q=1$ denote the differential binary phase-shift keying (DBPSK), $p=0.5$,  $p=0.5$ and $q=\log_2(M)/(8(M-1)^2)$ for  $M$-ary pulse amplitude modulation ($M$-PAM), and $p=0.5, q=0.125$ for non-return-to-zero (NRZ) on-off keyingon-off keying modulation.

To compare with the relay-assisted THz-RF link, we derive a closed-form expression of the BER $\bar{P}_{e1}$ for the direct THz link:
\begin{my_proposition}
	If $\phi$ and $S_0$ are the parameters of  pointing errors, and  $\alpha_1$ and $\mu_1$ are the fading parameters, then the average BER of the THz link is:
			
	\begin{flalign} \label{eq:pe_1_asy}
	&\bar{P}_{e1} =  \frac{A_1  C_1^{\hspace{-1mm}-\frac{\phi}{\alpha_1}} \Gamma(\mu_1)}{2\phi} \hspace{-0.5mm}+\hspace{-0.5mm} \frac{A_1  C_1^{\hspace{-1mm}-\frac{\phi}{\alpha_1}+1} \hspace{-1mm}{\gamma_1^0} ^{\frac{-\phi}{2}} \alpha_1^{\hspace{-1mm}\frac{\phi+2p-1}{2}}2^{B_1} q^{-\frac{\phi}{2}\hspace{-0.5mm}}}{2 \sqrt{2}\phi\Gamma(p)(2\pi)^{\alpha_1}}\nonumber\\& \times \hspace{-1mm} G_{2+\alpha_1,4}^{4,\alpha_1}\bigg(\begin{matrix} \Delta(\alpha_1,\alpha_1-\frac{\phi+2p}{2}), \Delta(2,1) \\ \Delta(2,B_1), \Delta(2,0)
	\end{matrix} \bigg|\frac{(C_1{\gamma_{1}^{0}}^{-\frac{\alpha_1}{2}}){\alpha_1}^{\alpha_1}}{4 q^{(\alpha_1)}}\bigg) \nonumber \\& - \frac{A_1 C_1^{-\frac{\phi}{\alpha_1}}{\gamma_1^0} ^{\frac{-\phi}{2}} 2^{(\mu_1-\frac{1}{2})}\alpha_1^{\frac{2p-1}{2}}}{2\phi\Gamma(p)(2\pi)^{\alpha_1}}\nonumber\\ &\times\hspace{-1mm} G_{2+\alpha_1,4}^{4,\alpha_1}\bigg(\hspace{-1mm}\begin{matrix} \Delta(\alpha_1,\alpha_1-p), \Delta(2,1) \\ \Delta(2,\mu_1), \Delta(2,0) \end{matrix} \bigg|\frac{(C_1{\gamma_{1}^{0}}^{-\frac{\alpha_1}{2}}){\alpha_1}^{\alpha_1}}{4 q^{(\alpha_1)}}\bigg) 
	\end{flalign}
\end{my_proposition}

\begin{IEEEproof}
	Using (\ref{eqn:cdf_thz}) in \eqref{eq:ber}, we get the average BER of the THz link as
	\begin{flalign} \label{eq:pe1_int_asy}
	\bar{P}_{e1} =& \frac{A_1  C_1^{-\frac{\phi}{\alpha_1}}q^p}{2\Gamma(p)\phi} \Bigg[ \int_{0}^{\infty} \gamma^{p-1} {e^{-q{\gamma}}} \Gamma(\mu_1)  d\gamma \nonumber \\ +&  \int_{0}^{\infty} C_1{\gamma_1^0} ^{\frac{-\phi}{2}}{\gamma} ^{\frac{\phi}{2}} \gamma^{p-1} {e^{-q{\gamma}}} \Gamma\Big(B_1,C_1{\gamma_1^0} ^{\frac{-\alpha_1}{2}}{\gamma} ^{\frac{\alpha_1}{2}}\Big) d\gamma \nonumber \\ -& \int_{0}^{\infty} \gamma^{p-1} {e^{-q{\gamma}}} \Gamma\Big(\mu_1,C_1{\gamma_1^0} ^{\frac{-\alpha_1}{2}}{\gamma} ^{\frac{\alpha_1}{2}}\Big)  d\gamma \Bigg]
	\end{flalign}
	
	The first integral in \eqref{eq:pe1_int_asy} contains a simple algebraic expression. For the second and third integrals, we apply the identity of definite integration of the product of two Meijer's G function \cite{Mathematica_two}. Thus, we get \eqref{eq:pe_1_asy}.
\end{IEEEproof}	 

For the sake of completeness, we use the standard procedure to present the average BER of the RF link $ (\bar{P}_{e2})$:
\begin{flalign} 
&\bar{P}_{e2} =  \frac{1}{2} -  \frac{2^{(\mu_2+\frac{1}{2})} \alpha_2^{\frac{2p-1}{2}}}{2\Gamma(p)\Gamma (\mu_2)(2\pi)^{(\alpha_2+1)}}\nonumber \\ \times &G_{4,\alpha_2}^{2+\alpha_2,4}\bigg(\begin{matrix} \Delta(\alpha_2,\alpha_2-p), \Delta(2,1) \\ \Delta(2,\mu_2), \Delta(2,0) \end{matrix} \bigg|\frac{(B_2{\gamma_{2}^{0}}^{-\frac{\alpha_2}{2}}){\alpha_2}^{\alpha_2}}{4 q^{(\alpha_2)}}\bigg) \label{eq:pe_2_asy}
\end{flalign}
Finally, we develop an analytical expression for the average BER of the relay-assisted system:
\begin{my_theorem} 
  Let $\phi$ and $S_0$ be the parameters of  pointing errors of the THz link. If  $\alpha_1$,  $\mu_1$ and  $\alpha_2$,  $\mu_2$ are  the fading parameters of THz  and RF links, respectively,  then an approximation of the average BER for the relay assisted THz-RF link is given as:
\begin{eqnarray}\label{eq_th_ber}
	\bar{P}_e\approx {\cal{F}}_{P_e}(p,q, \alpha_1, \alpha_2, \mu_1, \mu_2, \phi, S_0, \epsilon)
\end{eqnarray}
	where  ${\cal{F}}_{P_e}(p,q, \alpha_1, \alpha_2, \mu_1, \mu_2,\phi, S_0, \epsilon)$ is given in \eqref{eq:pe_3_asy}.
\end{my_theorem}	
\begin{IEEEproof}
	The proof is presented in Appendix E .
\end{IEEEproof}
It should be noted that the use of CDF to derive the average BER in \eqref{eq_th_ber} requires approximations on few terms only. Further, using $\alpha_1=\alpha_2=\alpha$, $\mu_1=\mu_2=\mu$, and $\epsilon=1$  in \eqref{eq_th_ber}, we can get an exact expression of the average BER for the i.i.d. fading model. It should be noted that the authors in \cite{Boulogeorgos_Error} have presented an expression for the average BER considering the i.i.d. model using the multi-variate Fox-H function.
\begin{my_corollary} \label{corr: ber}
	\begin{enumerate}[label=(\alph*)]
		\item  Considering the THz-link as Nakagami-$m$ fading  ($\alpha_1=2$, $\mu_1=2$) with  $\phi=2$, and the  Rayleigh fading ($\alpha_1=2$ $\mu_1=1$) for the RF-link,  the average BER of the relay assisted link is given as 
		\begin{flalign}\label{eq:special_ber1}
		\bar{P_e} = 1-\frac{1}{1+\gamma_2^0} + \frac{1}{2(1+ 2S_0^{-2}\gamma_1^0+\gamma_2^0)} - \frac{1+3\gamma_2^0}{(1+2\gamma_2^0)^2}
		\end{flalign} 		
		\item Considering the THz-link ($\alpha_1$$\to$$\infty$, $\mu_1=1$) as a linear Weibull fading with  $\phi=2$, and the  Rayleigh fading ($\alpha_2=2$ $\mu_2=1$) for the RF-link, the average BER of the relay assisted link: 		
		\begin{flalign}\label{eq:special_ber2}
		\bar{P_e} = \frac{\gamma_2^0 + 2 S_0^{-\phi}}{1+\gamma_2^0}
		\end{flalign} 					
	\end{enumerate}
\end{my_corollary}

\begin{IEEEproof}
 	Using $F(\gamma)$ from \eqref{cdf_relay} in \eqref{eq:ber} with special cases of $\alpha$ and $\mu$ enumerated in (a) and (b), and applying the standard procedure, we can get \eqref{eq:special_ber1} and \eqref{eq:special_ber2}.
\end{IEEEproof}
 
The results in Corollary \ref{corr: ber} simplifies the average BER performance for a few values of $\alpha$ and $\mu$ in order to provide insights on system behavior analytically. In the next section, we demonstrate the performance of relaying scheme for various values of $\alpha$, $\mu$, and other system parameters. 
\section{Simulation and Numerical Results}
\begin{table}[tp] 
	\caption{Parameters of the molecular absorption coefficient $k$ \cite{Boulogeorgos_Analytical}}	
	\label{table2}
	\begin{center}
		\begin{tabular}{|c|p {2cm}|c|p {2cm}|}
			\hline
			\textbf{Symbol}  & \centering{\textbf{Value}} & \textbf{Symbol}  & \textbf{Value}  \\
			\hline
			$q_1$ & $ 0.2205 $ & $q_2$ & $ 0.1303 $\\
			\hline
			$q_3$ & $ 0.0294 $ & $q_4$ & $ 0.4093 $\\
			\hline
			$q_5$ & $ 0.0925 $ & $q_6$ & $ 2.014 $\\
			\hline
			$q_7$ & $ 0.1702 $ & $q_8$ & $ 0.0303 $\\
			\hline
			$q_9$ & $ 0.537 $ & $q_{10}$ &  $ 0.0956 $\\
			\hline
			$c_1$ &  $5.54 \times 10^{-37}$ \mbox{$\rm Hz^{-3}$} & $c_4$ &  $-6.36\times10^{-3}$ \mbox{$\rm Hz^{-3}$}\\
			\hline
			$c_2$ &  $-3.94\times 10^{-25}$ \mbox{$\rm Hz^{-2}$} & $p_1$ &  $10.835$ \mbox{$\rm cm^{-1}$}\\
			\hline
			$c_3$ &  $9.06\times 10^{-14}$ \mbox{$\rm Hz^{-1}$} & $p_2$ &  $12.664$ \mbox{$\rm cm^{-1}$}\\
			\hline
		\end{tabular}
	\end{center}
\end{table}
\begin{table}[tp] 
	\caption{List of Simulation Parameters}
	\label{table3}	
	\begin{center}
		\begin{tabular}{ |p {4.2cm}| l | } 
			\hline
			\centering\textbf{Parameter} & \textbf{Value} \\ 
			\hline
			THz carrier frequency & $ 275 $ \mbox{GHz} \\ 
			\hline
			RF carrier frequency & $ 6 $ \mbox{GHz} \\ 
			\hline
			THz signal bandwidth & $ 10 $ \mbox{GHz} \\ 
			\hline
			RF signal bandwidth & $ 20 $ \mbox{MHz} \\ 			
			\hline
			Noise PSD & $ -174 $ \mbox{dBm/Hz} \\
			\hline
			Noise figure & $ 5 $ \mbox{dB} \\
			\hline
			Antenna Gain (dBi) & $ 55 $ (THz), $ 25 $ (RF) \\ 
			\hline
			$\alpha_1$, $\alpha_2$ & $1-6$ \\ 
			\hline
			$\mu_1$, $\mu_2$ & $0.5-4$ \\ 
			\hline
			$\Omega$ & $1$ \\ 
			\hline
				$p, q$ & $1,1$ (DBPSK) \\ 
			\hline
			Jitter standard deviation ($\sigma_s$) & $8-15$ \mbox{cm} \\ 
			\hline
			Antenna aperture radius $(r_1)$ & $ 10 $ \mbox{cm} \\ 
			\hline		
			\end{tabular}
	\end{center}
\end{table} 

In this section, we use numerical analysis and Monte Carlo simulations (averaged over $10^8$ channel realizations) to demonstrate the performance of the THz-RF relay assisted system. Although a direct link between the source and destination may not exist, we compare the performance of direct link with THz and relay-assisted transmissions for various scenarios.  We consider the THz link with a distance in the range of  $30-50$ \mbox{m}. This range is typical for the THz link, as adopted in \cite{faisal2020,Boulogeorgos_Error,Boulogeorgos_Analytical}. To compute the path loss for the THz link, we consider relative humidity, atmospheric pressure, and temperature as  $ 50\% $, \mbox{101325} Pa, and 296\textdegree K, respectively. The parameters for the calculation of the molecular absorption coefficient $k$ are provided in Table \ref{table2}. For parameters $S_0$ and $\phi$, we need radius  $r_1$ of the receiver antenna's effective aperture area $A_e$.  Using  $ A_e = \pi {r_1}^2 = {\lambda^2 G_1}/{4\pi} $ \cite{Balanis}, we can get $ r_1 = {\lambda\sqrt{G_1}}/{2\pi} $, where $G_1$ is the receiver antenna gain of the THz link. To simulate the pointing error, we consider the normalized beam-width $w_z/r_1$ in the range of $6$ to $12$.  The RF link distance is taken up to $50$ \mbox{m}, which is reasonable when users are connected to a nearby AP in a cell-free architecture. We compute the path loss of the RF link  using the 3GPP model $h_{l,2}({\rm dB}) = -(32.4+17.3\log_{10}(d_2)+20\log_{10} (10^{-9}f_2))$, where $d_2$ (in \mbox{m}) is the distance and $f_2$ (in \mbox{Hz}) is the carrier frequency of the RF link.  We use AWGN power  of  $-69.4$ \mbox{dBm} of the THz link over a bandwidth of $10$ \mbox{GHz} and  $-104.4$ \mbox{dBm} of the RF link over a bandwidth of $20$ \mbox{MHz} \cite{Sen_2020_Teranova}. Other simulation parameters for THz and RF systems as presented in Table \ref{table3}.
\begin{figure*}[tp]
	\begin{center}
		\subfigure[Different values of $\mu_1$ at $\alpha_1=2, \alpha_2=2,\mu_2=1, \sigma_s=8 {\rm cm},w_z/r_1=6, \gamma_{th}=4 \mbox{dB}$, $\phi=28.9576$, and  $S_0=0.054$.]{\includegraphics[width=\columnwidth]{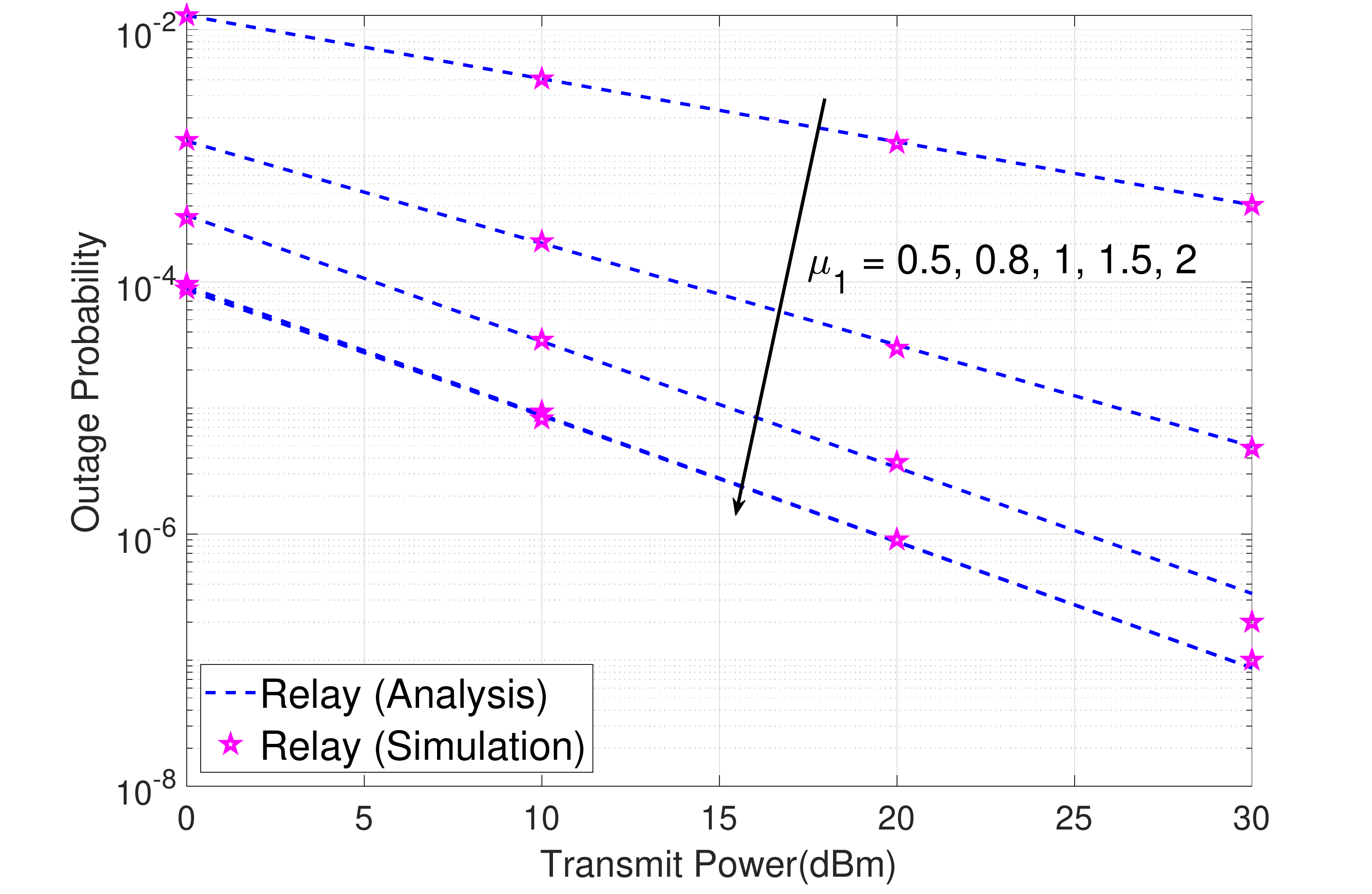}}
		\subfigure[Different values of $\alpha_1$ at $\mu_1=1, \alpha_2=2, \mu_2=4,\sigma_s=8 {\rm{cm}}, w_z/r_1=6, \gamma_{th}=4dB, \phi=28.9576$, and $S_0=0.054$.]{\includegraphics[width=\columnwidth]{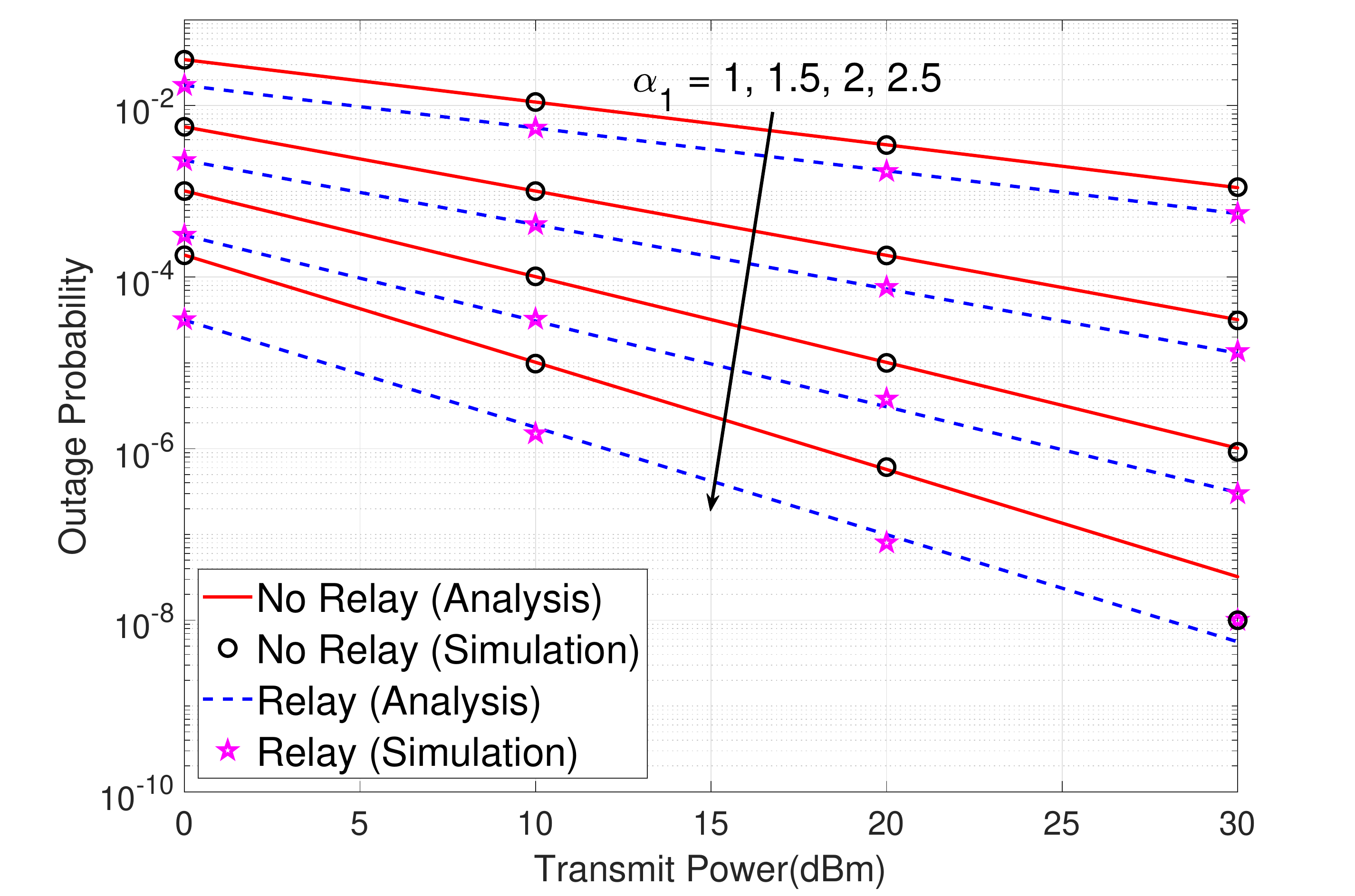}}
		\caption{Outage probability performance of relay-assisted THz-RF wireless link.}
		\label{outage}
	\end{center}
\end{figure*}
\begin{figure*}[tp]
	\begin{center}
		\subfigure[Different values of $w_z/r_1$  at $\alpha_1=2,\mu_1=4, \alpha_2=2,\mu_2=1$, and $\sigma_s=15 {\rm cm}$.]{\includegraphics[width=\columnwidth]{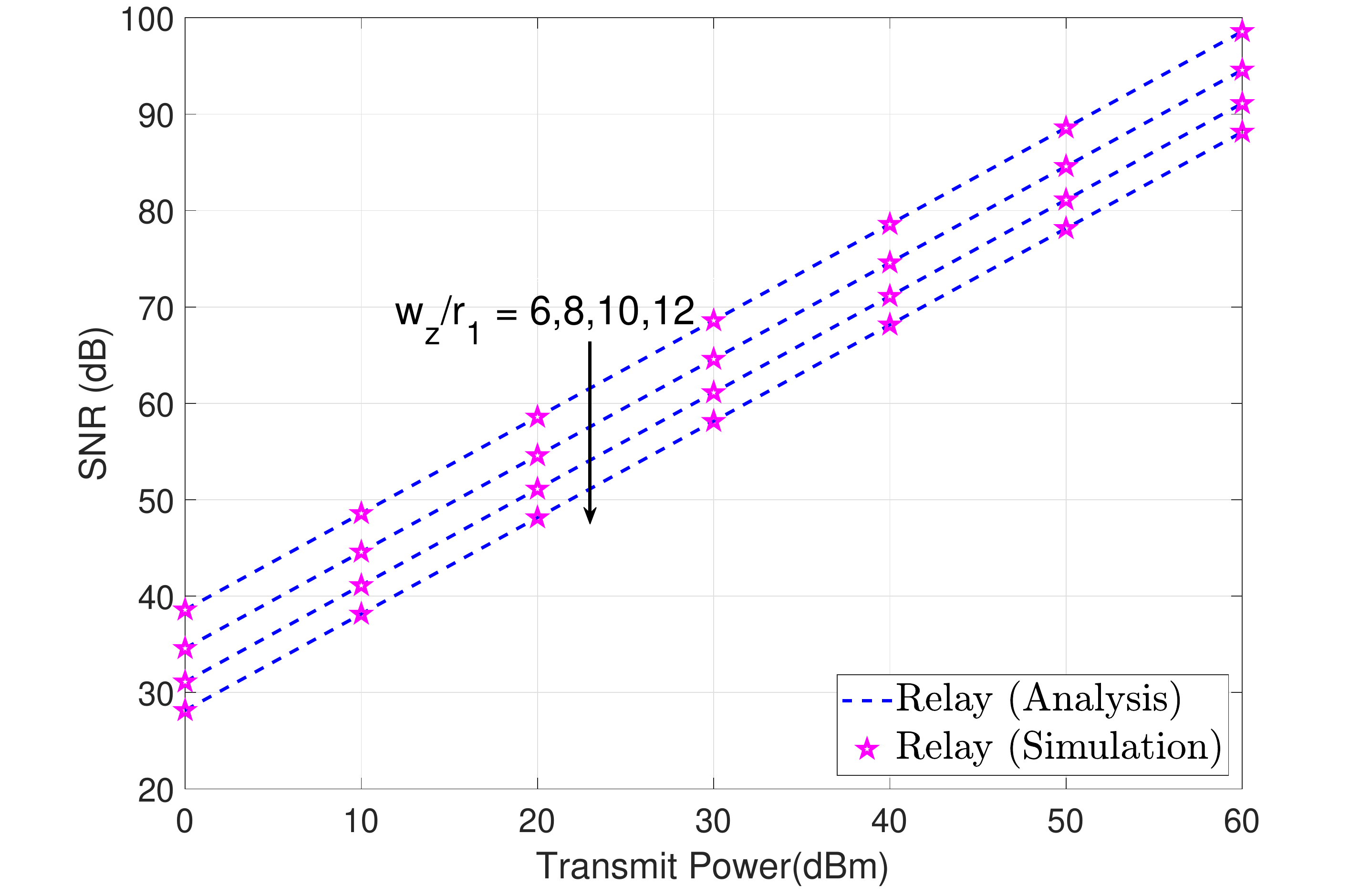}}
		\subfigure[Comparison between relayed and individual links at  $\alpha_1=2,\mu_1=4,\alpha_2=2, \mu_2=1,\sigma_s=15 {\rm cm}, w_z/r_1=6, \phi=8.2368$, and $S_0=0.054$.]{\includegraphics[width=\columnwidth]{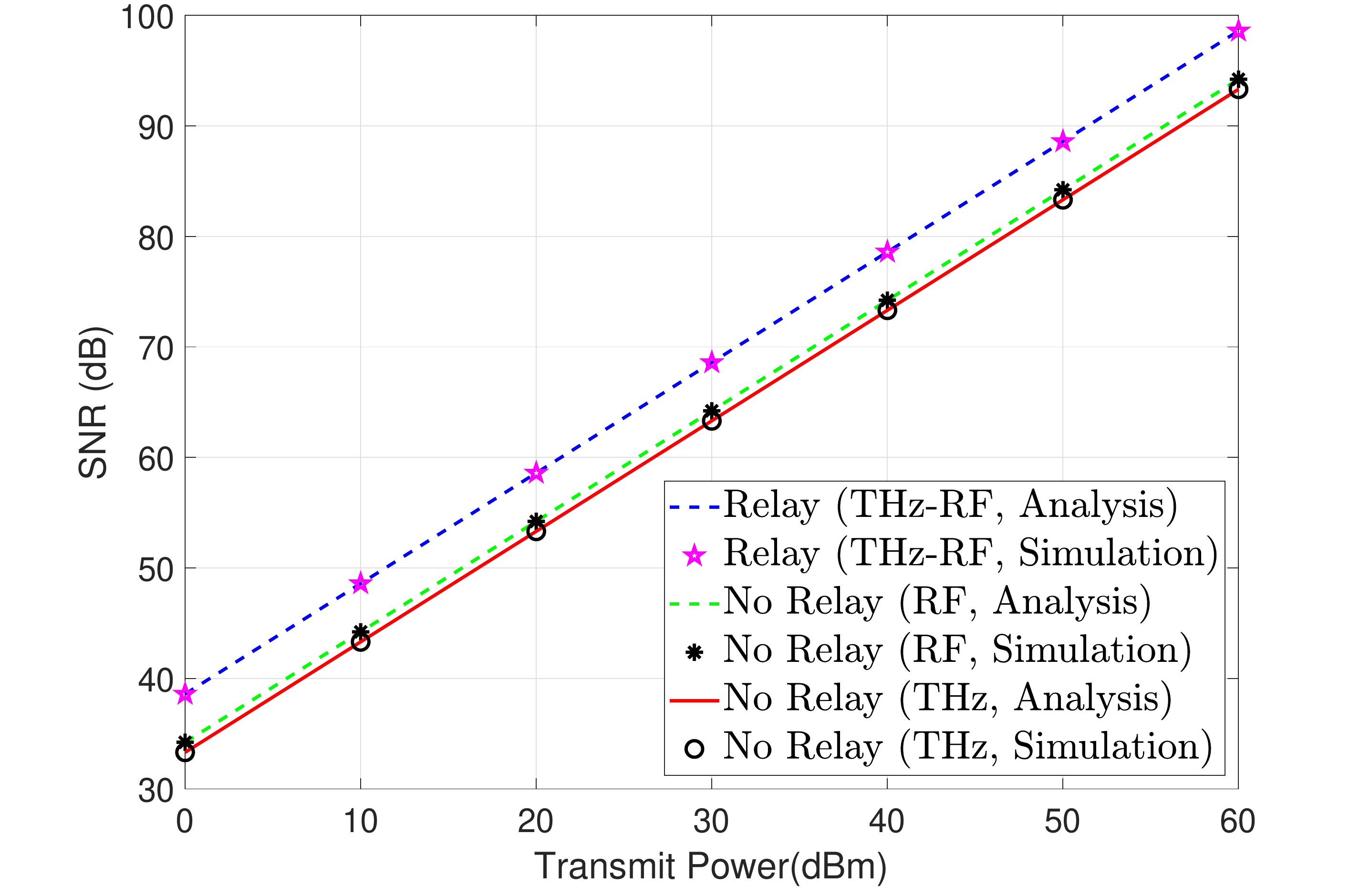}}
		\caption{Average SNR performance of relay-assisted THz-RF wireless link.}
		\label{snr}
	\end{center}
\end{figure*}

First, we demonstrate the outage probability performance of the relay assisted system, as shown in Fig.~\ref{outage}. In Fig.~\ref{outage}a, we show the impact of the parameter $\mu$  on the outage probability by  adopting the RF link as Rayleigh fading $\alpha_2=2$, $\mu_2=1$, and the THz link as the Nakagami-m fading $\alpha_1=2$ at different $\mu_1$. The outage performance improves with an increase in the parameter $\mu$ since an increase in $\mu_1$ accounts for dense clustering in the fading channel (i.e.,  good channel conditions). A very low value of $\mu_1<1$ (i.e., worse channel conditions, a typical scenario for the THz link) shows significant degradation in the outage performance. The derived CDF in Theorem \ref{th:cdf} allows the computation of the outage probability  for continuous (non-integer) values of $\mu_1$, which is necessary to evaluate the performance for a broader range of $\mu_1$, especially when $\mu_1<1$. It is noted that the state of art research use only integer values of $\mu_1$ to compute the CDF of the THz link \cite{Boulogeorgos_Error}. Further, we show the effect of the non-linearity parameter $\alpha_1$ on the outage probability.  In Fig.~\ref{outage}(b),  we consider $\alpha_2=2$, $\mu_2=4$ (Nakagami-m fading with higher clustering) for the RF link and  $\mu_1=2$ with different $\alpha_1$ of the THz link. The outage probability improves with an increase  in the parameter $\alpha_1$ (i.e., a decrease in the non-linearity of the THz fading). The outage probability dramatically improves with an increase in the parameter $\alpha_1$:  a factor of  $10^4$ decrease in the outage probability  when the parameter $\alpha_1$  increases from $1$ to $3$ at the same $10$ \mbox{dBm} of transmit power.

Observing Fig.~\ref{outage}a and Fig.~\ref{outage}b,  it can be emphasized that the diversity order  depends on $\alpha-\mu$ parameters of either of links since the  parameter $\phi$  is higher (low pointing errors) due to strong normalized beam-width $w_z=0.6$.   As such,  analysis in \eqref{diversity order}) shows that  the diversity order $M= \mu_1$ depends on the THz link when $\mu_1<1$, and  after that, there is no effect of the parameter $\mu_1$ since the  diversity order becomes $M=1$, which is confirmed in Fig.~\ref{outage}a. Similar conclusions on the outage probability behavior on the parameter $\alpha_1$ can be inferred from  Fig.~\ref{outage}b.

\begin{figure*}[tp]
	\begin{center}
		\subfigure[Different values of $w_z/r_1$  at $\alpha_1=2,\mu_1=4, \alpha_2=2,\mu_2=1$, and $\sigma_s=15 {\rm  cm}$.]{\includegraphics[width=\columnwidth]{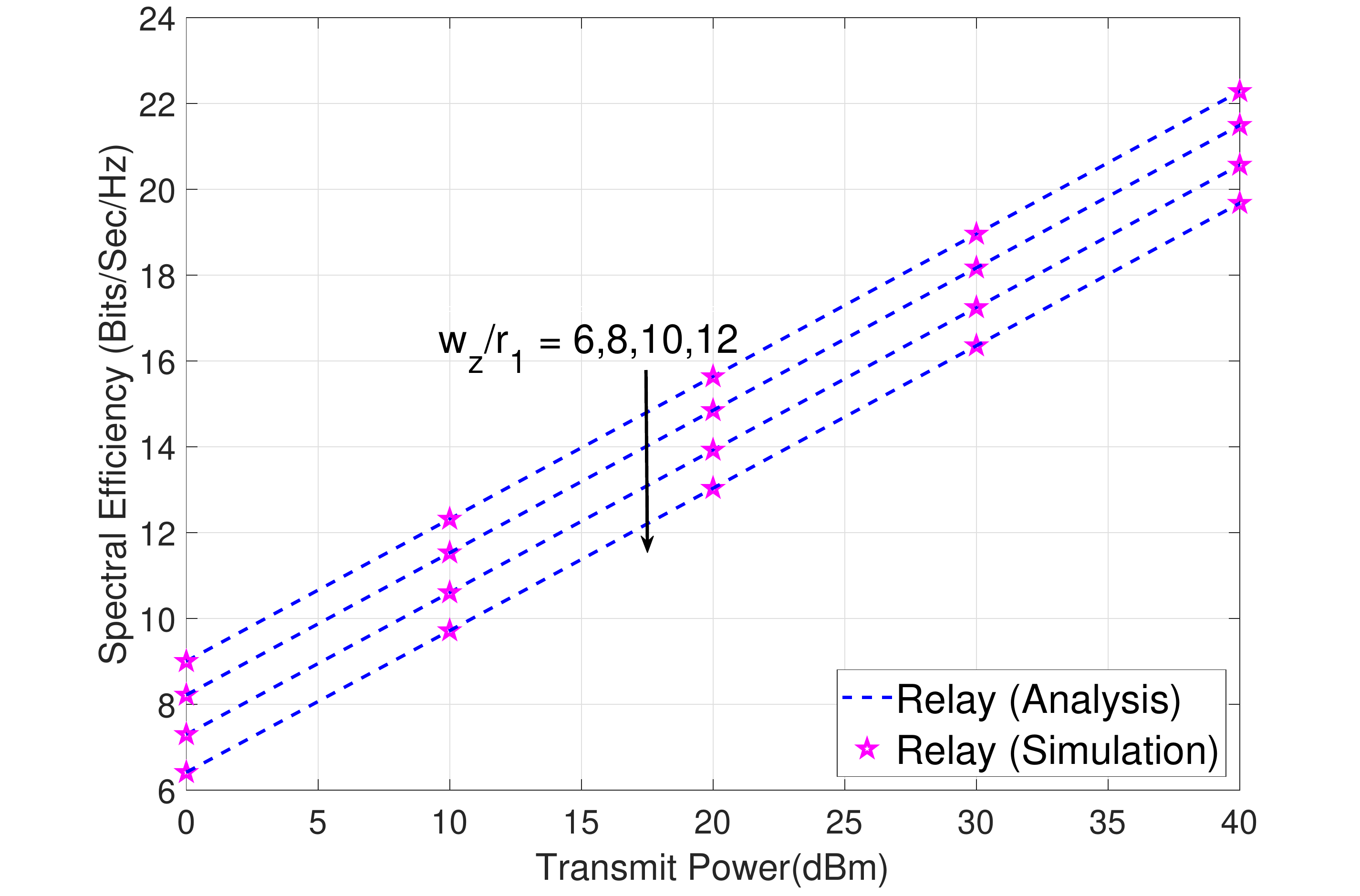}}
		\subfigure[Different values of $\alpha_1$, $\mu_1=1,\alpha_2=2,\mu_2=4,\sigma_s=15 {\rm cm}, w_z/r_1=6, \phi=8.2368$, and $S_0=0.054$.]{\includegraphics[width=\columnwidth]{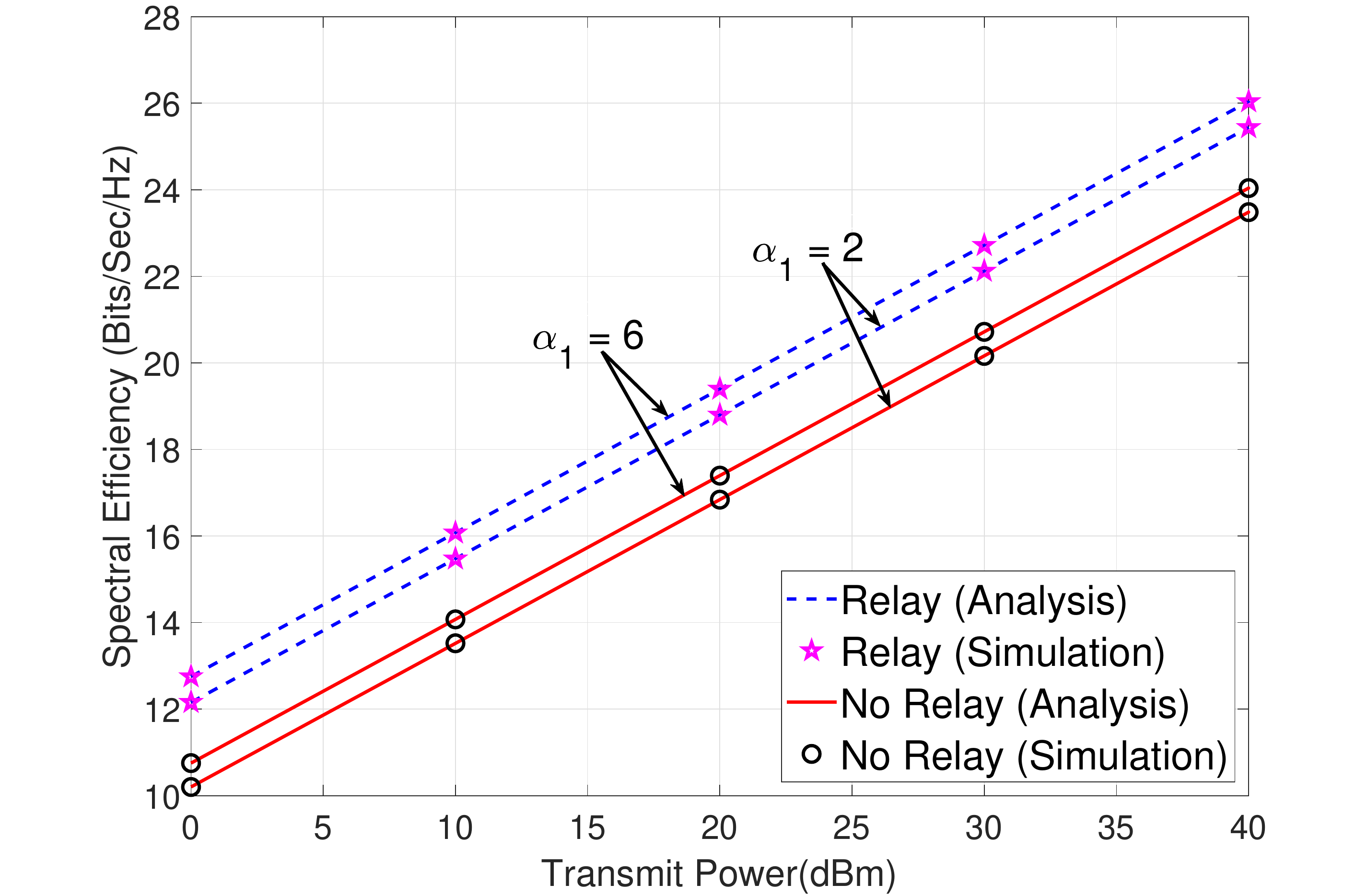}}
		\caption{Ergodic capacity performance of relay-assisted THz-RF wireless link.}
		\label{rate}
	\end{center}
\end{figure*}

Next, we  demonstrate the effect of pointing errors on the  average SNR and ergodic rate performance of the relay-assisted system, as shown  in Fig.~\ref{snr} and   Fig.~\ref{rate}. We  consider Rayleigh fading ($\alpha_2=2$, $\mu_2=1$) for the RF and Nakamani-m fading ($\alpha_1=2$, $\mu_1 =4$) for the THz link with    standard deviation of jitter ($\sigma_s=15$ \mbox{cm}) and different values of normalized beam-width $\frac{w_z}{r_1}$.  Fig.~\ref{snr}(a) and Fig.~\ref{rate}(a) demonstrate that the effect of pointing errors can be minimized by decreasing the normalized beam-width. It should be noted that the model of pointing errors in  	\eqref{eq:pdf_hp} is applicable when $w_z/r_1\geq 6$.  In Fig.~\ref{snr}(b),  we compare the performance of mixed THZ-RF  transmissions with direct RF and THz. It can be seen that the relay-assisted system provides a significant (around $ 5 $ \mbox{dB}) than the direct transmissions. There is a significant increase of $2$ \mbox{bits/sec/Hz} in the ergodic rate performance, as shown in Fig.~\ref{rate}(b). Although the direct link performance of THz and RF are almost similar, the THz has an enormous bandwidth (in the range of $10$ GHz) in comparison to the RF to get an enhanced channel capacity. It should be noted that the throughput of the THZ-RF system is limited by the bandwidth of RF link.
\begin{figure*}[tp]
	\begin{center}
		\subfigure[Different values of $\alpha_1$ and $w_z/r_1$ at $\mu_1=1, \alpha_2=2,\mu_2=4$, and $\sigma_s=15 {\rm cm}$.]{\includegraphics[width=\columnwidth]{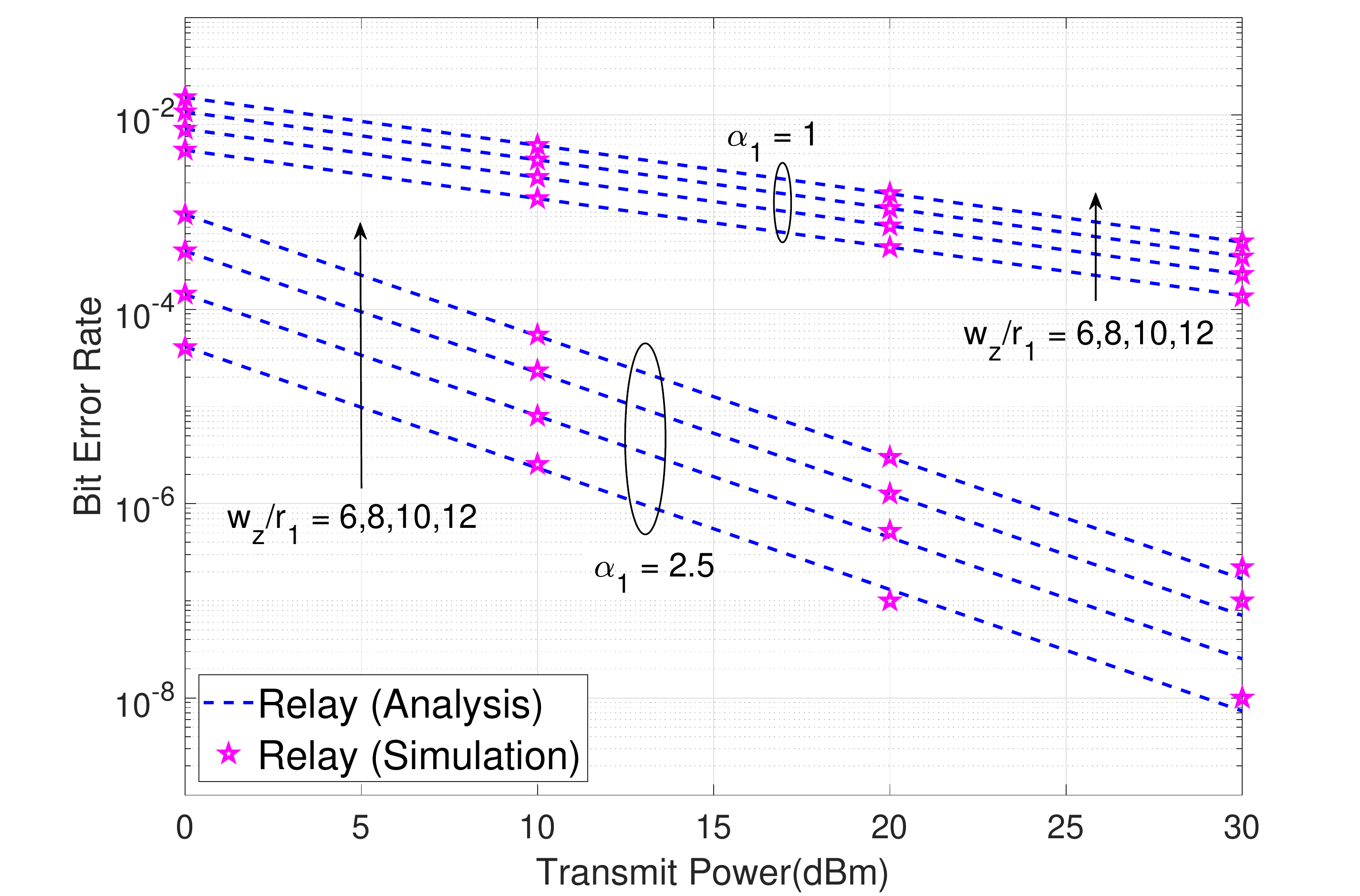}}
		\subfigure[Different link distances at $\alpha_1=2,\mu_1=1,\alpha_2=2,\mu_2=4,\sigma_s=15 {\rm cm}, w_z/r_1=6,\phi=8.2368$, and  $S_0=0.054$.]{\includegraphics[width=\columnwidth]{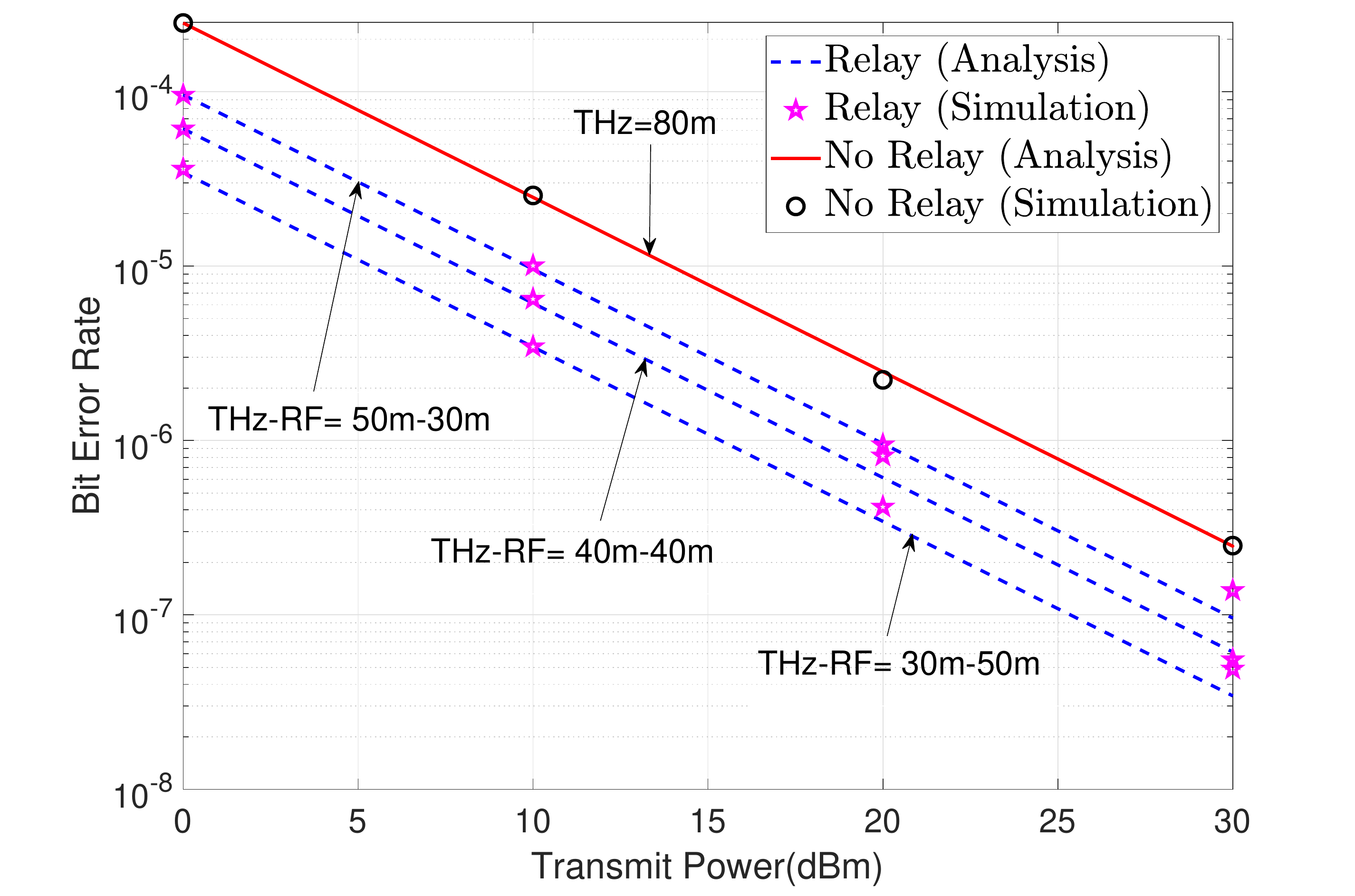}}
		\caption{Average BER performance of relay-assisted THz-RF wireless link.}
		\label{ber}
	\end{center}
\end{figure*}

Finally, we demonstrate the  average BER performance in Fig.~\ref{ber}. We show the effect of fading parameter $\alpha_1$ and normalized beam-wdith $w_z/r_1$ on the average BER in Fig.~\ref{ber}(a). The figure shows a dramatic improvement in the average BER with an  increase in $\alpha_1$ form $ 1 $ to $2.5$. However, the effect of normalized beam-width $w_z/r_1$ on the average BER is nominal.  In Fig.~\ref{ber}(b), we show the impact of relay location on the average BER performance for a link distance of $80$ \mbox{m}. It can be seen that a shorter THz link provides significant performance for the THz-RF relay system since the path loss at the THz frequencies is high.  
 
In all the above plots (Fig.~\ref{outage} to  Fig.~\ref{ber}), we  verified our derived expressions with the simulation and numerical results. It can be seen that the derived  analytical expressions of the outage probability and average SNR for both i.i.d and i.ni.d fading scenarios and average BER for the i.i.d. fading  have an exact match with the simulation results. The analytical results of ergodic capacity and average BER  for the i.ni.d fading can overestimate/underestimate the exact results due to  asymptotic bound $\Gamma(a,x)\to e^{-x}x^{a-1}$  in few terms of integration. However, Fig.~\ref{rate} and Fig.~\ref{ber} shows that the difference between analysis and simulation is indistinguishable.  Further, the analytical results of  ergodic capacity for the i.i.d fading are very close to the exact even with approximation $\log_2(1+x^a)\approx  a\log_2(1+x)$.  
 
\section{Conclusions}
We have investigated the performance of a THz-RF relay link over $\alpha$-$\mu$ fading with pointing errors. 
We considered a generalized  i.ni.d.  fading  model  by considering different $\alpha$ and $\mu$ parameters to model the short term fading  for both the THz and RF links, along with the statistical distribution of  pointing errors for  the THz link. By deriving a closed-form expression of the CDF for the THz link,  we analyzed the outage probability and diversity order of the system for real-valued $\mu$ and $\alpha$. The diversity order shows that the effect of pointing errors can be mitigated if the normalized beam width is adjusted sufficiently to get $\phi> \min \{\alpha_1 \mu_1, \alpha_2 \mu_2\}$. We  derived analytical expressions on average SNR, ergodic capacity, and average BER for a comprehensive analysis of the considered system. We also
develop simplified expressions to provide insight on the system behavior analytically under various practically relevant scenarios. Simulation and numerical analysis show a significant effect of fading parameters $\alpha$  and $\mu$ of the THz link on the THz-RF system.  However, compared with the fading parameters,  the effect of normalized beam-width on the performance of relay-assisted  is nominal.  We have shown a significant gain in the performance of the relay-assisted system compared with the direct transmissions. The THz-RF link can achieve higher data rates,  sufficient for data transmission between users and the central processing unit in a cell-free wireless network.  The proposed work can be augmented by investigating the THz-RF system using the constant gain AF relaying by deriving the PDF of SNR $\gamma= \frac{\gamma_1\gamma_2}{\gamma_2+C}$, where $C$ is a constant. 

\section* {Appendix A: Proof of Theorem 2}
Using (\ref{eqn:pdf_thz}), we define $\bar{\gamma}_1^{(n)}= \int_{0}^{\infty} \gamma^n f_1(\gamma) d\gamma$, and substituting $\big(\sqrt{{\gamma}/{\gamma_1^0}}\big)^{\alpha_1} = t$, we get the $n$-th moment of  SNR of the THz link
\begin{eqnarray} 
	\label{eq:gamma_1_int_asy}
	\bar{\gamma}_{1}^{(n)} = \frac{A_1 {\gamma_1^0}^n}{\alpha_1} \int_{0}^{\infty} t^{(\frac{\phi-\alpha_1+2n}{\alpha_1})} \Gamma (B_1,C_1t) dt 
\end{eqnarray}
Using the identity [\cite{DLMF}, 8.14.4] in  \eqref{eq:gamma_1_int_asy}, we solve the integral to get $\gamma_{1}^{(n)}$ of \eqref{eq:gamma_1_and_2_asy}. Similarly,  the $n$-th moment of  SNR of the RF link is $\bar{\gamma}_2^{(n)}$=$ \int_{0}^{\infty} \gamma^n f_2(\gamma) d\gamma$,  which can be solved easily to get \eqref{eq:gamma_1_and_2_asy}. Using \eqref{eqn:pdf_thz} and \eqref{eqn:cdf_rf} in  $\bar{\gamma}_{12}^{(n)}= \int_{0}^{\infty} \gamma^n f_1(\gamma) \gamma F_1(\gamma) d\gamma$, and substituting $\big(\sqrt{{\gamma}/{\gamma_1^0}}\big)^{\alpha_1} = t$, we get	

\begin{flalign} \label{eq:gamma_12_int_asy}
	&\bar{\gamma}_{12}^{(n)} = \frac{A_1 {\gamma_1^0}^n}{\alpha_1} \bigg[\int_{0}^{\infty} t^{(\frac{\phi-\alpha_1+2n}{\alpha_1})} ~\Gamma (B_1,C_1 t)~dt \nonumber \\ -& \frac{1}{\Gamma(\mu_2)}\int_{0}^{\infty} \hspace{-2mm}t^{(\frac{\phi-\alpha_1+2n}{\alpha_1})} \Gamma (B_1,C_1 t) {\Gamma\left(\mu_2, B_2' t^{\frac{\alpha_2}{\alpha_1}}\right)}dt\bigg] 
\end{flalign}
where $B_2' = B_2 \sqrt{{\gamma_1^0}/{\gamma_2^0}}$. The first integral in \eqref{eq:gamma_12_int_asy} is same as \eqref{eq:gamma_1_int_asy}.  To solve the second integral,  we use the  Meijer's G representation of $\Gamma (B_1,C_1 t)$ and $\Gamma\big(\mu_2, B_2' t^{\frac{\alpha_2}{\alpha_1}}\big) $ to get the second integral as
\begin{flalign}
	I_2 \hspace{-1mm}=\hspace{-1mm} \int_{0}^{\infty} \hspace{-2mm}t^{(\frac{\phi-\alpha_1+2n}{\alpha_1})} G_{1,2}^{2,0} \Big(\begin{matrix} 1 \\ B_1,0 \end{matrix} \Big|C_1 t \Big)  G_{1,2}^{2,0} \Big(\begin{matrix} 1 \\ \mu_2,0 \end{matrix} \Big| B_2't^{(\frac{\alpha_2}{\alpha_1})}\Big) dt
\end{flalign}
Using the identity \cite{Adamchik} of definite integration of product of two Meijer's G functions to get \eqref{eq:gamma_12_asy}. Similarly, using (\ref{eqn:pdf_rf}) and (\ref{eqn:cdf_thz}) in  $\bar{\gamma}_{21}^{(n)}= \int_{0}^{\infty} \gamma^n f_2(\gamma) \gamma F_1(\gamma) d\gamma$, and substituting $\big(\sqrt{{\gamma}/{\gamma_2^0}}\big)^\alpha = t$, we get	
\begin{flalign} \label{eq:gamma_21_int_asy}
	\bar{\gamma}_{21}^{(n)}&= \frac{A_1A_2C_1^{-\frac{\phi}{\alpha_1}}{\gamma_2^0}^n}{\Gamma(\mu_2)\phi} \bigg[ \int_{0}^{\infty} t^{(\mu_2-1+\frac{2n}{\alpha_2})} e^{(-B_2 t)} \Gamma(\mu_1) dt \nonumber \\ & +\int_{0}^{\infty} t^{(\mu_2-1+\frac{2n}{\alpha_2})} e^{(-B_2 t)} C_1' t^{\frac{\phi}{\alpha_2}}  \Gamma(B_1,C_1' t^{(\frac{\alpha_1}{\alpha_2})}) dt \nonumber \\ &-\int_{0}^{\infty} t^{(\mu_2-1+\frac{2n}{\alpha_2})} e^{(-B_2 t)} \Gamma(\mu_1,C_1't^{(\frac{\alpha_1}{\alpha_2})}) dt \bigg]
\end{flalign}
where $C_1' = C_1 \sqrt{{\gamma_2^0}/{\gamma_1^0}}$. The first integral in \eqref{eq:gamma_21_int_asy} is similar to $\bar{\gamma}_{2}^{(n)}$ and can be similarly  derived. To solve the second and third integrals, we use the identity \cite{Adamchik} of definite integration of the product of two Meijer's G functions to get \eqref{eq:gamma_21_asy}. 	
\section* {Appendix B: Proof of lemma 1}
Using (\ref{eqn:pdf_thz}), we define $\bar{\gamma}_1^{(n)}= \int_{0}^{\infty} \gamma^n f_1(\gamma) d\gamma$, and substituting $\big(\sqrt{{\gamma}/{\gamma_1^0}}\big)^\alpha = t$, we get the $n$-th moment of SNR of the THz link:
\begin{eqnarray} 
\label{eq:gamma_1_int}
\bar{\gamma}_{1}^{(n)} = \frac{A_1 {\gamma_1^0}^n}{\alpha} \int_{0}^{\infty} t^{(\frac{\phi-\alpha+2n}{\alpha})} \times \Gamma (B_1,C_1t) dt 
\end{eqnarray}
Using the identity [\cite{DLMF},8.14.4] in  \eqref{eq:gamma_1_int}, we solve the integral to get $\gamma_{1}^{(n)}$ of \eqref{eq:gamma_1_and_2}. Similarly,  the $n$-th moment of  SNR of the RF link is $\bar{\gamma}_2^{(n)}= \int_{0}^{\infty} \gamma^n f_2(\gamma) d\gamma$,  which can be solved easily  to  get \eqref{eq:gamma_1_and_2}. Using \eqref{eqn:pdf_thz} and \eqref{eqn:cdf_rf} in  $\bar{\gamma}_{12}^{(n)}= \int_{0}^{\infty} \gamma^n f_1(\gamma) \gamma F_1(\gamma) d\gamma$, and substituting $\big(\sqrt{{\gamma}/{\gamma_1^0}}\big)^\alpha = t$, we get
\begin{flalign} \label{eq:gamma_12_int}
\bar{\gamma}_{12}^{(n)} &= \frac{A_1 {\gamma_1^0}^n}{\alpha} \bigg[\int_{0}^{\infty} t^{(\frac{\phi-\alpha+2n}{\alpha})} \Gamma (B_1,C_1 t)dt \nonumber \\ &- \int_{0}^{\infty} t^{(\frac{\phi-\alpha+2n}{\alpha})} \Gamma (B_1,C_1 t) \left(\frac{\Gamma\left(\mu, B_2' t\right)}{\Gamma (\mu)}\right)dt\bigg] 
\end{flalign}
where $B_2' = B_2 \sqrt{{\gamma_1^0}/{\gamma_2^0}}$. The first integral in \eqref{eq:gamma_12_int} is the  same as \eqref{eq:gamma_1_int}.  To solve the second integral,  we apply the integration by parts tasking  $\Gamma\left(\mu, B_2' t\right)$ as the first term and $t^{(\frac{\phi-\alpha+2n}{\alpha})} \Gamma (B_1,C_1 t)$ as the second term, and use the identity  [\cite{Gradshteyn},eq.(6.455/1)] to get \eqref{eq:gamma_12}. Similarly, using (\ref{eqn:pdf_rf}) and (\ref{eqn:cdf_thz}) in  $\bar{\gamma}_{21}^{(n)}= \int_{0}^{\infty} \gamma^n f_2(\gamma) \gamma F_1(\gamma) d\gamma$, and substituting $\big(\sqrt{{\gamma}/{\gamma_2^0}}\big)^\alpha = t$, we get

\begin{flalign} \label{eq:gamma_21_int}
\bar{\gamma}_{21}^{(n)}&= \frac{A_1A_2C_1^{-\frac{\phi}{\alpha}}{\gamma_2^0}^n}{\Gamma(\mu)\phi} \bigg[ \int_{0}^{\infty} t^{(\mu-1+\frac{2n}{\alpha})} e^{(-B_2 t)} \Gamma(\mu) dt \nonumber \\ & +\int_{0}^{\infty} t^{(\mu-1+\frac{2n}{\alpha})} e^{(-B_2 t)} C_1' t^{\frac{\phi}{\alpha}}  \Gamma(B_1,C_1' t) dt \nonumber \\ &-\int_{0}^{\infty} t^{(\mu-1+\frac{2n}{\alpha})} e^{(-B_2 t)} \Gamma(\mu,C_1't) dt \bigg]
\end{flalign}
where $C_1' = C_1 \sqrt{{\gamma_2^0}/{\gamma_1^0}}$. The first integral in \eqref{eq:gamma_21_int} is similar to $\bar{\gamma}_{2}^{(n)}$ and can be similarly derived. For the second and third integrals, we use the identity 
 [\cite{Gradshteyn}, eq.(6.455/1)] to get \eqref{eq:gamma_21}. 
\section* {Appendix C: Proof of Theorem 3}
Using \eqref{cdf_relay} in  \eqref{eq:total capacity_cdf}, we represent the ergodic capacity of the relay assisted THz-RF link:
\begin{flalign} \label{eq:eta_12_int_asy}
	&\bar{\eta} = \frac{1}{\Gamma(\mu_2)\rm ln(2)} \int_{0}^{\infty} \frac{1}{1+\gamma} {\Gamma\left(\mu_2, B_2 {\gamma_2^{0}}^{\frac{-\alpha_2}{2}} \gamma^{\frac{\alpha_2}{2}}\right)} d\gamma  \nonumber \\ -& \frac{A_1  C_1^{-\frac{\phi}{\alpha_1}} }{\phi \Gamma(\mu_2)\rm ln(2)} \bigg[  \int_{0}^{\infty} \frac{1}{1+\gamma}  \Gamma(\mu_1) {\Gamma\left(\mu_2, B_2 {\gamma_2^{0}}^{\frac{-\alpha_2}{2}} \gamma^{\frac{\alpha_2}{2}}\right)}  d\gamma  \nonumber \\   \nonumber \\ -& \hspace{-1mm} \int_{0}^{\infty} \hspace{-3mm} \frac{\gamma^{\frac{\phi}{2}}}{1+\gamma}  C_1 {\gamma_1^{0}}^{\frac{-\phi}{2}}  \Gamma\big(B_1,C_1 {\gamma_1^{0}}^{\frac{-\alpha_1}{2}} \hspace{-1mm}\gamma^{\frac{\alpha_1}{2}} \big) {\Gamma \hspace{-1mm}\left(\hspace{-1mm}\mu_2, B_2 {\gamma_2^{0}}^{\frac{-\alpha_2}{2}} \hspace{-1mm}\gamma^{\frac{\alpha_2}{2}}\hspace{-1mm}\right)}  d\gamma  \nonumber \\ +& \int_{0}^{\infty} \frac{1}{1+\gamma}  \Gamma\big(\mu_1,C_1 {\gamma_1^{0}}^{\frac{-\alpha_1}{2}} \gamma^{\frac{\alpha_1}{2}}\big) {\Gamma\left(\mu_2, B_2 {\gamma_2^{0}}^{\frac{-\alpha_2}{2}} \gamma^{\frac{\alpha_2}{2}}\right)}    \bigg] d\gamma 
\end{flalign}
	
We solve the first and second integral in \eqref{eq:eta_12_int_asy} exactly by applying the identity of definite integration of product of one power and one Meijer's G function \cite{Mathematica}. For the third integral, we use the series expansion of  ${\Gamma\big(\mu_2, B_2 {\gamma_2^{0}}^{\frac{-\alpha_2}{2}} \gamma^{\frac{\alpha_2}{2}}\big)}$ using $\Gamma(a,bx) = (a-1)! \exp(-bx) \sum_{k=0}^{a-1}\frac{(bx)^k}{k!}$ and asymptotic expansion of    $\Gamma\big(B_1,C_1 {\gamma_1^{0}}^{\frac{-\alpha_1}{2}} \gamma^{\frac{\alpha_1}{2}} \big) $ using $\Gamma(a,x) \approx  e^{-x}x^{a-1}$. Employing  $\alpha_2 = \epsilon \alpha_1$,  the third  integral can be represented as  	
\begin{flalign}\label{eq3rd}
	\bar{\eta}_{I_3}\hspace{-1mm} \approx &\hspace{-2mm} \sum_{K_2=0}^{\mu_2-1} \hspace{-2mm} \frac{A_1  C_1^{-\frac{\phi}{\alpha_1}+1} \hspace{-1mm}{\gamma_1^0}^{\frac{-\phi}{2}} \hspace{-1mm}(B_2 {\gamma_{2}^0}^{-\frac{\alpha_2}{2}}\hspace{-1mm})^{k_2} (e^{\hspace{-1mm}-B_2{\gamma_{2}^0}^{-\frac{\alpha_2}{2}}}\hspace{-1mm}) (C_1{\gamma_1^0} ^{\frac{-\alpha_1}{2}}\hspace{-1mm})^{B_1-1}}{\phi \rm{ln(2)} k_2!} \nonumber \\ \times & \int_{0}^{\infty} \hspace{-2mm}{\gamma} ^{\frac{\phi+K_2\alpha_2+B_1 \alpha_1 -3}{2}}   e^{-\gamma^\frac{\alpha_1}{2}(C_1{\gamma_1^0}^{\frac{-\alpha_1}{2}}+ \gamma^{\frac{\alpha_1(\epsilon-1)}{2}})} d\gamma
\end{flalign}
Thus, we apply the identity of definite integration of the product of one power and one  Meijer's G function \cite{Mathematica} to solve $\bar{\eta}_{I_3}$ in \eqref{eq3rd}. For the fourth integral, we use the series expansion for gamma functions. Similar to the $\bar{\eta}_{I_3}$, we use  $\alpha_2 = \epsilon \alpha_1$ and apply the identity of definite integration of the product of one power and one  Meijer's G function \cite{Mathematica} to solve the fourth integral. Capitalizing these, we get the ergodic capacity of THz-RF relay system  in \eqref{eq:eta_asy} of Theorem 3. It should be noted that the approximation is used in the last two terms that are not significant compared to the  first two terms. 
\section* {Appendix D: Proof of Lemma 2}
Using (\ref{eqn:pdf_thz}) in \eqref{eq:total capacity_pdf}, we define $\bar{\eta}_1 \geq \int_{0}^{\infty} {\log}_2(\gamma) f_1(\gamma)d\gamma$, and substituting $\big(\sqrt{{\gamma}/{\gamma_1^{0}}}\big)^\alpha = t$, we get a lower bound  average capacity of THz link:	

\begin{equation} \label{eq:eta_1_int}
\bar{\eta}_1\geq\frac{A_1}{\alpha  \log(2)}\int_{0}^{\infty} {\rm log} (\gamma_1^0 t^{\frac{2}{\alpha}}) t^{(\frac{\phi}{\alpha}-1)}  \Gamma (B_1,C_1 t) dt 
\end{equation}	
It is noted that $\eta_1$ is a lower bound on the ergodic capacity of the THz link since $\log_2(1+\gamma)\geq \log_2(\gamma)$. 
To find a closed-form expression,  we use integration by-parts with $\Gamma (B_1,C_1 t)$ as the first term and $ {\rm log} (\gamma_1^0 t^{\frac{2}{\alpha}}) t^{(\frac{\phi}{\alpha}-1)}$ as the second term,  and apply the identity [\cite{Gradshteyn}, (eq.4.352/1)] to get $\bar{\eta}_1$ in \eqref{eq:eta_1_asy}. Similarly, substituting $\big(\sqrt{{\gamma}/{\gamma_2^0}}\big)^\alpha=t$ in  $\bar{\eta}_2\geq \int_{0}^{\infty} \rm{log}_2(\gamma) f_2(\gamma)d\gamma$ for a lower bound on the average capacity of the RF link:
\begin{equation} \label{eq:eta_2_int}
\bar{\eta}_2 \geq \frac{A_2}{\Gamma(\mu) \rm log(2)}\int_{0}^{\infty} { \log} (\gamma_2^0 t^{\frac{2}{\alpha}}) t^{(\mu-1)} e^{(-B_2 t)} dt
\end{equation}
We use the identity [\cite{Gradshteyn}, (eq.4.352/1)] in \eqref{eq:eta_2_int} to get $\bar{\eta}_2$ of \eqref{eq:eta_2_asy}.
Defining $\bar{\eta}_{12} = \int_{0}^{\infty} {{\log}}_2(\gamma) f_1(\gamma) F_2(\gamma)d\gamma$, using \eqref{eqn:pdf_thz} and \eqref{eqn:cdf_rf}, and substituting $\big(\sqrt{{\gamma}/{\gamma_1^0}}\big)^\alpha = t$, we get
\begin{eqnarray} \label{eq:eta_12_int}
\bar{\eta}_{12} \geq \frac{A_1}{\alpha \rm log(2)} \bigg[\int_{0}^{\infty} {\rm log} (\gamma_1^0t^{\frac{2}{\alpha}}) t^{(\frac{\phi}{\alpha}-1)} \Gamma (B_1,C_1 t) dt \nonumber \\ - \int_{0}^{\infty} {\rm log} (\gamma_1^0 t^{\frac{2}{\alpha}}) t^{(\frac{\phi}{\alpha}-1)} \Gamma (B_1,C_1 t) \left(\frac{\Gamma\left(\mu, B_2' t\right)}{\Gamma (\mu)}\right) dt \bigg]  
\end{eqnarray}
The first integral in \eqref{eq:eta_12_int} is the same as \eqref{eq:eta_1_int}. To solve the second integral,  we use the series expansion of Gamma function $\Gamma(\mu,B_2't) = (\mu-1)! \exp(-B_2't) \sum_{k=0}^{\mu-1}\frac{(B_2't)^k}{k!}$.  Using $ {\rm log} (\gamma_1^0 t^{\frac{2}{\alpha}}) \approx \frac{2}{\alpha} {\rm log} (1+{\gamma_1^0}^{\frac{\alpha}{2}}t) $, and with  Meijer's G representation of ${\rm log} (1+{\gamma_1^0}^{\frac{\alpha}{2}}t)$, $\Gamma (B_1,C_1 t)$ and $\Gamma\left(\mu, B_2' t\right) $, we get the second integral as
\begin{flalign}
I_2 &\approx \sum_{k=0}^{\mu-1} \frac{2A_1(\mu-1)! B_2'^k}{\alpha^2 \rm log(2) \Gamma(\mu) k!} \int_{0}^{\infty} t^{(\frac{\phi}{\alpha}+k-1)} G_{0,1}^{1,0} \Big(\begin{matrix} - \\ 0 \end{matrix} \Big|B_2' t \Big)  \nonumber \\ &\times G_{1,2}^{2,0} \Big(\begin{matrix} 1 \\ B_1,0 \end{matrix} \Big|C_1 t \Big)  \Big[G_{2,2}^{1,2}\Big(\begin{matrix} 1,1 \\ 1,0 \end{matrix} \Big|(\gamma_1^0)^{\frac{\alpha}{2}} t \Big)\Big] dt
\end{flalign}
Finally, we apply the identity of definite integration of the product of three Meijer's G function \cite{Mathematica_three} to get $\bar{\eta}_{12}$ in \eqref{eq:eta_12}. 
Similarly, using \eqref{eqn:pdf_rf} and \eqref{eqn:cdf_thz} we define $\bar{\eta}_{21}$ $\geq$ $\int_{0}^{\infty} {\log}_2(\gamma)f_2(\gamma) F_1(\gamma)d\gamma$ and substituting $\sqrt{{\gamma}/{\gamma_2^0}}^\alpha$=$t$, we get
\begin{flalign}  \label{eq:eta_21_int}
\bar{\eta}_{21} &\approx \frac{A_1  C_1^{-\frac{\phi}{\alpha}} A_2}{\Gamma(\mu)\phi {\rm log(2)}} \bigg[\int_{0}^{\infty} {\rm log} ( \gamma_2^0 t^{\frac{2}{\alpha}}) t^{(\mu-1)} e^{(-B_2 t)}  \Gamma(\mu)dt  \nonumber \\ &+ \int_{0}^{\infty} {\rm log} ( \gamma_2^0 t^{\frac{2}{\alpha}}) t^{(\mu-1)} e^{(-B_2 t)} C_1' t^{\frac{\phi}{\alpha}}  \Gamma(B_1,C_1't)dt \nonumber \\ &- \int_{0}^{\infty} {\rm log} ( \gamma_2^0 t^{\frac{2}{\alpha}}) t^{(\mu-1)} e^{(-B_2 t)} \Gamma(\mu,C_1't)dt \bigg] 
\end{flalign}
The first integral in \eqref{eq:eta_21_int}  is similar to $\eta_2$ of \eqref{eq:eta_2_int}. To solve the second integral, we apply the identity of definite integration of the product of three Meijer's G function \cite{Mathematica_three}. Finally, to solve the third integration, we use the series expansion of Gamma function $\Gamma(\mu,C_1't) = (\mu-1)! \exp(-C_1't) \sum_{k=0}^{\mu-1}\frac{(C_1't)^k}{k!}$ and apply the identity of definite integration of the product of two Meijer's G function \cite{Mathematica_two}. Combining these three integrations, we get $\bar{\eta}_{21}$ in \eqref{eq:eta_21}.

\section* {Appendix E: Proof of Theorem 4}
Using \eqref{cdf_relay} in  \eqref{eq:ber}, the average BER of the relay assisted THz-RF link can be expressed as

 \begin{eqnarray} \label{eq:pe3_int_asy}
	&\bar{P}_{e} =\hspace{-1mm} \frac{A_1  C_1^{-\frac{\phi}{\alpha_1}}q^p}{2 \Gamma(p)\phi \Gamma(\mu_2)} \Bigg[\hspace{-1mm} \int_{0}^{\infty} \hspace{-3mm} \gamma^{p-1} {e^{-q{\gamma}}} \Gamma(\mu_1 \hspace{-1mm}) {\Gamma\big(\mu_2, B_2 {\gamma_2^0} ^{\frac{-\alpha_2}{2}}\hspace{-1mm}{\gamma} ^{\frac{\alpha_2}{2}} \hspace{-1mm}\big)} d\gamma \nonumber \\  &+ {C_1{\gamma_1^0} ^{\frac{-\phi}{2}}} \hspace{-1mm}\int_{0}^{\infty} \hspace{-3mm} {\gamma} ^{\frac{\phi+2p-2}{2}} {e^{ \hspace{-1mm}- \hspace{-0.5mm}q{\gamma}}} \Gamma\Big( \hspace{-1mm}B_1, \hspace{-1mm}C_1{\gamma_1^0} ^{\frac{-\alpha_1}{2}} \hspace{-1mm}{\gamma} ^{\frac{\alpha_1}{2}} \hspace{-1mm}\Big) {\Gamma\big(\mu_2, B_2 {\gamma_2^0} ^{\frac{-\alpha_2}{2}} \hspace{-1mm}{\gamma} ^{\frac{\alpha_2}{2}} \hspace{-1mm}\big)} d\gamma \nonumber \\ &- \hspace{-1mm} \int_{0}^{\infty} \hspace{-2mm} \gamma^{p-1} {e^{ \hspace{-1mm}- \hspace{-0.5mm}q{\gamma}}} \Gamma\Big(\mu_1,C_1{\gamma_1^0} ^{\frac{-\alpha_1}{2}} \hspace{-1mm}{\gamma} ^{\frac{\alpha_1}{2}} \hspace{-1mm}\Big)  {\Gamma\big(\mu_2, B_2 {\gamma_2^0} ^{\frac{-\alpha_2}{2}} \hspace{-1mm}{\gamma} ^{\frac{\alpha_2}{2}} \hspace{-1mm}\big)} d\gamma \Bigg] \nonumber \\ &+\frac{q^p}{2\Gamma(p)}\int_{0}^{\infty} \gamma^{p-1} {e^{-q{\gamma}}} \bigg(1-\frac{\Gamma\big(\mu_2, B_2 {\gamma_2^0} ^{\frac{-\alpha_2}{2}}{\gamma} ^{\frac{\alpha_2}{2}}\big)}{\Gamma (\mu_2)}\bigg)  d\gamma	
\end{eqnarray}
To solve the first integral in \eqref{eq:pe3_int_asy}, we apply the identity of definite integration of product of two Meijer's G function \cite{Mathematica_two}. For the second integral, we use the series expansion for ${\Gamma\big(\mu_2, B_2 {\gamma_2^{0}}^{\frac{-\alpha_2}{2}} \gamma^{\frac{\alpha_2}{2}}\big)}$ using $\Gamma(a,bx)$ =$(a-1)! e^{(-bx)}$ $\sum_{k=0}^{a-1}\frac{(bx)^k}{k!}$ and   approximation for  $\Gamma\big(B_1,C_1 {\gamma_1^{0}}^{\frac{-\alpha_1}{2}} \gamma^{\frac{\alpha_1}{2}} \big) $ using $\Gamma(a,x) \approx e^{-x}x^{a-1}$. Finally, using $\alpha_2 = \epsilon \alpha_1$, the second integral becomes   
\begin{flalign}\label{eqberlast}
	&\bar{P}_{eI_2}  \hspace{-1mm}\approx\hspace{-2mm} \sum_{K_2=0}^{\mu_2-1} \hspace{-2mm} \frac{A_1  C_1^{ \hspace{-1mm}-\frac{\phi}{\alpha_1}+1} \hspace{-1mm}{\gamma_1^0} ^{\hspace{-0.5mm}\frac{-\phi}{2}} \hspace{-1mm} (\hspace{-0.5mm}B_2 {\gamma_{2}^0}^{\hspace{-0.5mm}-\frac{\alpha_2}{2}}\hspace{-1mm})^{\hspace{-0.5mm}k_2} ( \hspace{-0.5mm}e^{ \hspace{-1mm}-B_2{\gamma_{2}^0}^{-\frac{\alpha_2}{2}}} \hspace{-1mm}) (\hspace{-0.5mm}C_1{\gamma_1^0} ^{\hspace{-0.5mm}\frac{-\alpha_1}{2}} \hspace{-1mm})^{B_1 \hspace{-0.5mm}-1}q^p}{2\Gamma(p)\phi k_2!} \nonumber \\ &\times  \hspace{-2mm}\int_{0}^{\infty} \hspace{-3mm}{\gamma} ^{\frac{\phi+K_2\alpha_2+ \alpha_1(B_1-1)}{2}} \hspace{-1mm}e^{p-1} {e^{\hspace{-1mm}-q{\gamma}}}   e^{\hspace{-1mm}-\gamma^{\hspace{-1mm}\frac{\alpha_1}{2}}\hspace{-0.5mm}(C_1{\gamma_1^0}^{\frac{-\alpha_1}{2}}\hspace{-1mm}+ \gamma^{\hspace{-1mm}\frac{\alpha_1(\epsilon-1)}{2}}\hspace{-0.5mm})} d\gamma
\end{flalign}
Employing the identity of definite integration of the product of two Meijer's G function \cite{Mathematica_two},  we solve the integral in \eqref{eqberlast}. For the third integral in \eqref{eq:pe3_int_asy}, we use the series expansion for  gamma functions, and $\alpha_2 = \epsilon \alpha_1$  to get an expression of the third integral. The fourth integral in \eqref{eq:pe3_int_asy} can be solved applying the identity of definite integration of the product of two Meijer's G function \cite{Mathematica_two}. Capitalizing the expressions of these four integral, we get the average BER for the relay assisted link  in \eqref{eq_th_ber} of Theorem 4.

\bibliographystyle{ieeetran}
\bibliography{thz_bib_file}

\begin{thebibliography}{10}
\providecommand{\url}[1]{#1}
\csname url@samestyle\endcsname
\providecommand{\newblock}{\relax}
\providecommand{\bibinfo}[2]{#2}
\providecommand{\BIBentrySTDinterwordspacing}{\spaceskip=0pt\relax}
\providecommand{\BIBentryALTinterwordstretchfactor}{4}
\providecommand{\BIBentryALTinterwordspacing}{\spaceskip=\fontdimen2\font plus
\BIBentryALTinterwordstretchfactor\fontdimen3\font minus
  \fontdimen4\font\relax}
\providecommand{\BIBforeignlanguage}[2]{{%
\expandafter\ifx\csname l@#1\endcsname\relax
\typeout{** WARNING: IEEEtran.bst: No hyphenation pattern has been}%
\typeout{** loaded for the language `#1'. Using the pattern for}%
\typeout{** the default language instead.}%
\else
\language=\csname l@#1\endcsname
\fi
#2}}
\providecommand{\BIBdecl}{\relax}
\BIBdecl

\bibitem{pranay2021}
P.~Bhardwaj and {{S.M.} Zafaruddin}, ``Performance of dual-hop relaying for
  {THz-RF} wireless link,'' \emph{To be presented at 2021 IEEE 93rd Vehicular
  Technology Conference (VTC2021-Spring), Helsinki, Finland, 25-28 April 2021,
  preprint version: arXiv:2020.12186}, 2020.

\bibitem{Bhushan2014densification}
N.~{Bhushan}, J.~{Li}, D.~{Malladi}, R.~{Gilmore}, D.~{Brenner},
  A.~{Damnjanovic}, R.~T. {Sukhavasi}, C.~{Patel}, and S.~{Geirhofer},
  ``Network densification: the dominant theme for wireless evolution into
  {5G},'' \emph{IEEE Communications Magazine}, vol.~52, no.~2, pp. 82--89,
  2014.

\bibitem{Liu2017densification}
J.~{Liu}, M.~{Sheng}, L.~{Liu}, and J.~{Li}, ``Network densification in {5G}:
  From the short-range communications perspective,'' \emph{IEEE Communications
  Magazine}, vol.~55, no.~12, pp. 96--102, 2017.

\bibitem{Dulaimi2017densification}
A.~{Al-Dulaimi}, S.~{Al-Rubaye}, J.~{Cosmas}, and A.~{Anpalagan}, ``Planning of
  ultra-dense wireless networks,'' \emph{IEEE Network}, vol.~31, no.~2, pp.
  90--96, 2017.

\bibitem{Koenig_2013_nature}
S.~{Koenig}, D.{Lopez-Diaz}, and J.{Antes et al.}, ``Wireless {sub-THz}
  communication system with high data rate,'' \emph{Nature Photon}, vol.~7, p.
  977–981, 2013.

\bibitem{Elayan_2019}
H.~{Elayan}, O.~{Amin}, B.~{Shihada}, R.~M. {Shubair}, and M.~{Alouini},
  ``Terahertz band: The last piece of {RF} spectrum puzzle for communication
  systems,'' \emph{IEEE Open Journal of the Communications Society}, vol.~1,
  pp. 1--32, 2020.

\bibitem{faisal2020}
A.~{Faisal}, H.~{Sarieddeen}, H.~{Dahrouj}, T.~Y. {Al-Naffouri}, and M.~S.
  {Alouini}, ``Ultramassive {MIMO} systems at {Terahertz} bands: Prospects and
  challenges,'' \emph{IEEE Vehicular Technology Magazine}, vol.~15, no.~4, pp.
  33--42, 2020.

\bibitem{Wang2015}
N.~{Wang}, E.~{Hossain}, and V.~K. {Bhargava}, ``Backhauling {5G} small cells:
  A radio resource management perspective,'' \emph{IEEE Wireless
  Communications}, vol.~22, no.~5, pp. 41--49, 2015.

\bibitem{Wang2014}
C.~{Wang}, B.~{Lu}, C.~{Lin}, Q.~{Chen}, L.~{Miao}, X.~{Deng}, and J.~{Zhang},
  ``{0.34-THz} wireless link based on high-order modulation for future wireless
  local area network applications,'' \emph{IEEE Transactions on Terahertz
  Science and Technology}, vol.~4, no.~1, pp. 75--85, 2014.

\bibitem{Boulogeorgos_Analytical}
A.~A. {Boulogeorgos} and A.~{Alexiou}, ``Analytical performance assessment of
  {THz} wireless systems,'' \emph{IEEE Access}, vol.~7, pp. 11\,436--11\,453,
  2019.

\bibitem{Boulogeorgos_Error}
------, ``Error analysis of mixed {THz-RF} wireless systems,'' \emph{IEEE
  Communications Letters}, vol.~24, no.~2, pp. 277--281, 2020.

\bibitem{Jornet_2011}
J.~M. {Jornet} and I.~F. {Akyildiz}, ``Channel modeling and capacity analysis
  for electromagnetic wireless nanonetworks in the {Terahertz} band,''
  \emph{IEEE Transactions on Wireless Communications}, vol.~10, no.~10, pp.
  3211--3221, 2011.

\bibitem{Priebe_2011}
S.~{Priebe}, C.~{Jastrow}, M.~{Jacob}, T.~{Kleine-Ostmann}, T.~{Schrader}, and
  T.~{Kürner}, ``Channel and propagation measurements at 300 {GHz},''
  \emph{IEEE Transactions on Antennas and Propagation}, vol.~59, no.~5, pp.
  1688--1698, 2011.

\bibitem{Kim2015}
S.~{Kim} and A.~G. {Zajić}, ``Statistical characterization of {300-GHz}
  propagation on a desktop,'' \emph{IEEE Transactions on Vehicular Technology},
  vol.~64, no.~8, pp. 3330--3338, 2015.

\bibitem{Kokkoniemi_2018}
J.~{Kokkoniemi}, J.~{Lehtomäki}, and M.~{Juntti}, ``Simplified molecular
  absorption loss model for 275–400 {Gigahertz} frequency band,'' in
  \emph{12th European Conference on Antennas and Propagation (EuCAP 2018)},
  2018, pp. 1--5.

\bibitem{Wu2020}
Y.~{Wu}, J.~{Kokkoniemi}, C.~{Han}, and M.~{Juntti}, ``Interference and
  coverage analysis for {Terahertz} networks with indoor blockage effects and
  line-of-sight access point association,'' \emph{IEEE Transactions on Wireless
  Communications}, vol.~20, no.~3, pp. 1472--1486, 2021.

\bibitem{Sarieddeen2019}
H.~{Sarieddeen}, M.~{Alouini}, and T.~Y. {Al-Naffouri}, ``{Terahertz}-band
  ultra-massive spatial modulation {MIMO},'' \emph{IEEE Journal on Selected
  Areas in Communications}, vol.~37, no.~9, pp. 2040--2052, 2019.

\bibitem{Cheng_2020}
C.~{Cheng}, S.~{Sangodoyin}, and A.~{Zajić}, ``{Terahertz MIMO} fading
  analysis and doppler modeling in a data center environment,'' in \emph{2020
  14th European Conference on Antennas and Propagation (EuCAP)}, 2020, pp.
  1--5.

\bibitem{Olutayo_2020}
A.~{Olutayo}, J.~{Cheng}, and J.~F. {Holzman}, ``A new statistical channel
  model for emerging wireless communication systems,'' \emph{IEEE Open Journal
  of the Communications Society}, vol.~1, pp. 916--926, 2020.

\bibitem{Bian2021}
J.~{Bian}, C.~X. {Wang}, X.~{Gao}, X.~{You}, and M.~{Zhang}, ``A general {3D}
  non-stationary wireless channel model for {5G} and beyond,'' \emph{IEEE
  Transactions on Wireless Communications}, pp. 1--1, 2021.

\bibitem{boronin2015}
P.~{Boronin}, D.~{Moltchanov}, and Y.~{Koucheryavy}, ``A molecular noise model
  for {THz} channels,'' in \emph{2015 IEEE International Conference on
  Communications (ICC)}, 2015, pp. 1286--1291.

\bibitem{Petrov2015}
V.~{Petrov}, D.~{Moltchanov}, and Y.~{Koucheryavy}, ``Interference and {SINR}
  in dense {Terahertz} networks,'' in \emph{2015 IEEE 82nd Vehicular Technology
  Conference (VTC2015-Fall)}, 2015, pp. 1--5.

\bibitem{zhang2016}
R.~{Zhang}, K.~{Yang}, A.~{Alomainy}, Q.~H. {Abbasi}, K.~{Qaraqe}, and R.~M.
  {Shubair}, ``Modelling of the {Terahertz} communication channel for in-vivo
  nano-networks in the presence of noise,'' in \emph{2016 16th Mediterranean
  Microwave Symposium (MMS)}, 2016, pp. 1--4.

\bibitem{Elayan2018noisemodel}
H.~{Elayan}, C.~{Stefanini}, R.~M. {Shubair}, and J.~M. {Jornet}, ``End-to-end
  noise model for intra-body {Terahertz} nanoscale communication,'' \emph{IEEE
  Transactions on NanoBioscience}, vol.~17, no.~4, pp. 464--473, 2018.

\bibitem{Boluda_2017}
R.~{Boluda-Ruiz}, A.~{García-Zambrana}, C.~{Castillo-Vázquez},
  B.~{Castillo-Vázquez}, and S.~{Hranilovic}, ``Outage performance of
  exponentiated {Weibull FSO} links under generalized pointing errors,''
  \emph{Journal of Lightwave Technology}, vol.~35, no.~9, pp. 1605--1613, 2017.

\bibitem{KOKKONIEMI2020}
J.~{Kokkoniemi}, A.~{Boulogeorgos}, M.~{Aminu}, J.~{Lehtomäki}, A.~{Alexiou},
  and M.~{Juntti}, ``Impact of beam misalignment on {THz} wireless systems,''
  \emph{Nano Communication Networks}, vol.~24, p. 100302, 2020.

\bibitem{Rong_2017}
Z.~{Rong}, M.~S. {Leeson}, and M.~D. {Higgins}, ``Relay-assisted nanoscale
  communication in the {THz} band,'' \emph{Micro Nano Letters}, vol.~12, no.~6,
  pp. 373--376, 2017.

\bibitem{Abbasi_2017}
Q.~H. {Abbasi}, A.~A. {Nasir}, K.~{Yang}, K.~A. {Qaraqe}, and A.~{Alomainy},
  ``Cooperative {In-Vivo} nano-network communication at {Terahertz}
  frequencies,'' \emph{IEEE Access}, vol.~5, pp. 8642--8647, 2017.

\bibitem{Boulogeorgos_20020_THz_Relaying}
A.~A. {Boulogeorgos}, E.~N. {Papasotiriou}, and A.~{Alexiou}, ``Outage
  probability analysis of {THz} relaying systems,'' \emph{arXiv:2007.07186},
  2020.

\bibitem{Mir2020}
T.~{Mir}, M.~{Waqas}, U.~{Mir}, S.~M. {Hussain}, A.~M. {Elbir}, and S.~{Tu},
  ``Hybrid precoding design for two-way relay-assisted {Terahertz} massive
  {MIMO} systems,'' \emph{IEEE Access}, vol.~8, pp. 222\,660--222\,671, 2020.

\bibitem{Farid2007}
A.~A. {Farid} and S.~{Hranilovic}, ``Outage capacity optimization for
  free-space optical links with pointing errors,'' \emph{Journal of Lightwave
  Technology}, vol.~25, no.~7, pp. 1702--1710, 2007.

\bibitem{Yacoub_alpha_mu}
M.~D. {Yacoub}, ``The $\alpha$-$\mu$ distribution: A physical fading model for
  the stacy distribution,'' \emph{IEEE Transactions on Vehicular Technology},
  vol.~56, no.~1, pp. 27--34, 2007.

\bibitem{RIS_THz_HW_Impaiment}
H.~Du, J.~Zhang, K.~Guan, B.~Ai, and T.~Kürner, ``Reconfigurable intelligent
  surface aided terahertz communications under misalignment and hardware
  impairments,'' \emph{arXiv:2012.00267}, 2020.

\bibitem{Nosratinia2004}
A.~{Nosratinia}, T.~E. {Hunter}, and A.~{Hedayat}, ``Cooperative communication
  in wireless networks,'' \emph{IEEE Communications Magazine}, vol.~42, no.~10,
  pp. 74--80, 2004.

\bibitem{Li2012}
Q.~{Li}, R.~Q. {Hu}, Y.~{Qian}, and G.~{Wu}, ``Cooperative communications for
  wireless networks: techniques and applications in {LTE}-advanced systems,''
  \emph{IEEE Wireless Communications}, vol.~19, no.~2, pp. 22--29, 2012.

\bibitem{Bjornson_2013}
E.~{Bjornson}, M.~{Matthaiou}, and M.~{Debbah}, ``A new look at dual-hop
  relaying: Performance limits with hardware impairments,'' \emph{IEEE
  Transactions on Communications}, vol.~61, no.~11, pp. 4512--4525, 2013.

\bibitem{Lee2011}
E.~{Lee}, J.~{Park}, D.~{Han}, and G.~{Yoon}, ``Performance analysis of the
  asymmetric dual-hop relay transmission with mixed {RF/FSO} links,''
  \emph{IEEE Photonics Technology Letters}, vol.~23, no.~21, pp. 1642--1644,
  2011.

\bibitem{Ansari2013}
I.~S. {Ansari}, F.~{Yilmaz}, and M.~{Alouini}, ``Impact of pointing errors on
  the performance of mixed {RF/FSO} dual-hop transmission systems,'' \emph{IEEE
  Wireless Communications Letters}, vol.~2, no.~3, pp. 351--354, 2013.

\bibitem{Samimi2013}
H.~{Samimi} and M.~{Uysal}, ``End-to-end performance of mixed {RF/FSO}
  transmission systems,'' \emph{IEEE/OSA Journal of Optical Communications and
  Networking}, vol.~5, no.~11, pp. 1139--1144, 2013.

\bibitem{assym_rf_fso2015}
S.~{Anees} and M.~R. {Bhatnagar}, ``Performance of an amplify and forward dual
  hop asymmetric {RF FSO} communication system,'' \emph{IEEE/OSA Journal of
  Optical Communications and Networking}, vol.~7, no.~2, pp. 124--135, February
  2015.

\bibitem{series_hybrid_m_channel2015}
L.~{Kong}, W.~{Xu}, L.~{Hanzo}, H.~{Zhang}, and C.~{Zhao}, ``Performance of a
  free-space-optical relay-assisted hybrid {RF/FSO} system in generalized
  {$M$}-distributed channels,'' \emph{IEEE Photonics Journal}, vol.~7, no.~5,
  pp. 1--19, Oct 2015.

\bibitem{dual_hop_rf_fso_turb2016}
E.~{Zedini}, H.~{Soury}, and M.~{Alouini}, ``On the performance analysis of
  dual-hop mixed {FSO/RF} systems,'' \emph{IEEE Transactions on Wireless
  Communications}, vol.~15, no.~5, pp. 3679--3689, May 2016.

\bibitem{Bag2018}
B.~{Bag}, A.~{Das}, I.~S. {Ansari}, A.~{Prokeš}, C.~{Bose}, and A.~{Chandra},
  ``Performance analysis of hybrid {FSO} systems using {FSO/RF-FSO} link
  adaptation,'' \emph{IEEE Photonics Journal}, vol.~10, no.~3, pp. 1--17, 2018.

\bibitem{Zhang2020}
Y.~{Zhang}, J.~{Zhang}, L.~{Yang}, B.~{Ai}, and M.~{Alouini}, ``On the
  performance of dual-hop systems over mixed {FSO/mmWave} fading channels,''
  \emph{IEEE Open Journal of the Communications Society}, vol.~1, pp. 477--489,
  2020.

\bibitem{Boulogeorgos_performance_2018}
A.~A. {Boulogeorgos}, E.~N. {Papasotiriou}, J.~{Kokkoniemi}, J.~{Lehtomaeki},
  A.~{Alexiou}, and M.~{Juntti}, ``Performance evaluation of {THz} wireless
  systems operating in 275-400 {GHz} band,'' in \emph{2018 IEEE 87th Vehicular
  Technology Conference (VTC Spring)}, 2018, pp. 1--5.

\bibitem{papoulis_2002}
{A. Papoulis} and {S. Pillai}, \emph{Probability, Random Variables, and
  Stochastic Processes}.\hskip 1em plus 0.5em minus 0.4em\relax McGraw Hill,
  Boston, Fourth Edition, 2002.

\bibitem{Jameson2016}
G.~J.~O. {Jameson}, ``The incomplete gamma functions,'' \emph{The Mathematical
  Gazette}, vol. 100, Iss. 548, pp. 298--306, Jul 2016.

\bibitem{Annamalai2010}
A.~{Annamalai}, R.~C. {Palat}, and J.~{Matyjas}, ``Estimating ergodic capacity
  of cooperative analog relaying under different adaptive source transmission
  techniques,'' in \emph{2010 IEEE Sarnoff Symposium}, 2010, pp. 1--5.

\bibitem{Gradshteyn}
I.~S. {Gradshteyn} and I.~M. {Ryzhik }, \emph{Table of Integrals, Series, and
  Products}.\hskip 1em plus 0.5em minus 0.4em\relax Academic press, San Diego,
  CA, 6th edition, 2000.

\bibitem{Ansari2011}
I.~S. {Ansari}, S.~{Al-Ahmadi}, F.~{Yilmaz}, M.~{Alouini}, and
  H.~{Yanikomeroglu}, ``A new formula for the {BER} of binary modulations with
  dual-branch selection over generalized-k composite fading channels,''
  \emph{IEEE Transactions on Communications}, vol.~59, no.~10, pp. 2654--2658,
  2011.

\bibitem{Mathematica_two}
\emph{The Wolfram function Site}, Accessed: Sept. 26, 2020. Available:
  http://functions.wolfram.com/HypergeometricFunctions/MeijerG/21/02/03/01.

\bibitem{Balanis}
C.~{Balanis}, \emph{Antenna Theory: Analysis and Design}.\hskip 1em plus 0.5em
  minus 0.4em\relax John Wiley and Sons, 3rd Edition, 2016.

\bibitem{Sen_2020_Teranova}
P.~{ Sen}, D.~A. {Pados}, S.~N. {Batalama}, E.~{Einarsson}, J.~P. {Bird}, and
  J.~M. {Jornet}, ``The {TeraNova} platform: An integrated testbed for
  ultra-broadband wireless communications at true {Terahertz} frequencies,''
  \emph{Computer Networks, Elsevier}, 2020.

\bibitem{DLMF}
\emph{Incomplete {Gamma} and Realted Functions}, Accessed: Sept. 26, 2020
  [Online]. Available: http://dlmf.nist.gov/8.14.

\bibitem{Adamchik}
V.~{Adamchik} and O.~{Marichev}, \emph{The algorithm for calculating integrals
  of hypergeometric type functions and its realization in reduce system}.\hskip
  1em plus 0.5em minus 0.4em\relax John Wiley and Sons, 3rd Edition, 1990.

\bibitem{Mathematica}
\emph{The Wolfram function Site}, Accessed: Jan. 08, 2021. Available:
  https://functions.wolfram.com/HypergeometricFunctions/MeijerG/21/02/07/.

\bibitem{Mathematica_three}
\emph{The Wolfram function Site}, Accessed: Sept. 26, 2020. Available:
  http://functions.wolfram.com/07.34.21.0012.01.

\end{thebibliography}
\end{document}